\documentclass[structabstract,longauth]{aa}
\usepackage[utf8]{inputenc}
\usepackage{threeparttable, pifont}
\usepackage{natbib, graphicx}
\usepackage{pstricks, color, multicol}
\usepackage{amsmath, amssymb, url}
\usepackage{hyperref}
\usepackage{caption,subcaption}
\usepackage{txfonts, longtable, dcolumn, verbatim, xcolor}
\bibpunct{(}{)}{;}{a}{}{,}
\usepackage[normalem]{ulem}
\usepackage[squaren,textstyle]{SIunits}

\newcommand{\adam}{\texttt{ADAM}}

\newcommand{\mistral}{\texttt{Mistral}}

\newcommand{\Diam}{332\,$\pm$\,6}
\newcommand{\Mass}{(3.79\,$\pm$\,1.28)\,$\times$\,$10^{19}$}
\newcommand{\Masss}{3.79\,$\pm$\,1.28}
\newcommand{\Dens}{1.98\,$\pm$\,0.68}
\newcommand{\sid}{g$\cdot$cm$^{-3}$}
\newcommand{\sidd}{kg$\cdot$m$^{-3}$}
\begin{document}

\title{ {(704)~Interamnia: A transitional object between a dwarf planet and a typical irregular-shaped minor body}
    \thanks{Based on observations made with ESO Telescopes at the La Silla Paranal Observatory under program 199.C-0074 (PI Vernazza)},\thanks{The reduced  images  are  available  at  the  CDS  via  anonymous   ftp  to \url{http://cdsarc.u-strasbg.fr/} or via \url{http://cdsarc.u-strasbg.fr/viz-bin/qcat?J/A+A/xxx/Axxx}}}
\titlerunning{Physical model of (704)~Interamnia}

\author{
    J.~Hanu{\v s}\inst{\ref{prague}}       \and 
    P.~Vernazza\inst{\ref{lam}}            \and 
    M.~Viikinkoski\inst{\ref{tampere}}     \and 
    M.~Ferrais\inst{\ref{liege},\ref{lam}}           \and 
    N.~Rambaux\inst{\ref{imcce}}           \and 
    E.~Podlewska-Gaca\inst{\ref{poznan}}   \and 
    A.~Drouard\inst{\ref{lam}}             \and 
    L.~Jorda\inst{\ref{lam}}               \and 
    E.~Jehin\inst{\ref{liege}}             \and 
    B.~Carry\inst{\ref{oca}}               \and 
    M.~Marsset\inst{\ref{mit}}             \and 
    F.~Marchis\inst{\ref{seti}}            \and 
    B.~Warner\inst{\ref{warner}}           \and 
    R.~Behrend\inst{\ref{behrend}}         \and 
    V.~Asenjo\inst{\ref{asenjo}}           \and 
    N.~Berger\inst{\ref{berger}}           \and 
    M.~Bronikowska\inst{\ref{poznan2}}     \and 
    T.~Brothers\inst{\ref{mit}}            \and 
    S.~Charbonnel\inst{\ref{charbonnel}}   \and 
    C.~Colazo\inst{\ref{cordoba}}          \and 
    J-F.~Coliac\inst{\ref{coliac}}         \and 
    R.~Duffard\inst{\ref{duffard}}         \and 
    A.~Jones\inst{\ref{jones}}             \and 
    A.~Leroy\inst{\ref{leroy}}             \and 
    A.~Marciniak\inst{\ref{poznan}}        \and 
    R.~Melia\inst{\ref{cordoba}}           \and 
    D.~Molina\inst{\ref{molina}}           \and 
    J.~Nadolny\inst{\ref{tenerife}}        \and 
    M.~Person\inst{\ref{mit}}              \and 
    O.~Pejcha\inst{\ref{teorka}}           \and 
    H.~Riemis\inst{\ref{riemis}}           \and 
    B.~Shappee\inst{\ref{hawaii}}          \and 
    K.~Sobkowiak\inst{\ref{poznan}}        \and 
    F.~Sold\'an\inst{\ref{soldan}}         \and 
    D.~Suys\inst{\ref{riemis}}             \and 
    R.~Szakats\inst{\ref{konkoly}}         \and 
    J.~Vantomme\inst{\ref{riemis}}         \and 
    M.~Birlan\inst{\ref{imcce}}            \and 
    J.~Berthier\inst{\ref{imcce}}          \and 
    P.~Bartczak\inst{\ref{poznan}}         \and 
    C.~Dumas\inst{\ref{tmt}}               \and 
    G.~Dudzi\'{n}ski\inst{\ref{poznan}}    \and
    J.~{\v D}urech\inst{\ref{prague}}      \and 
    J.~Castillo-Rogez\inst{\ref{jpl}}      \and 
    F.~Cipriani\inst{\ref{estec}}          \and 
    R.~Fetick\inst{\ref{lam}}              \and 
    T.~Fusco\inst{\ref{lam}}               \and 
    J.~Grice\inst{\ref{oca},\ref{ou}}      \and 
    M.~Kaasalainen\inst{\ref{tampere}}     \and 
    A.~Kryszczynska\inst{\ref{poznan}}     \and 
    P.~Lamy\inst{\ref{lamos}}              \and 
    T.~Michalowski\inst{\ref{poznan}}      \and 
    P.~Michel\inst{\ref{oca}}              \and 
    T.~Santana-Ros\inst{\ref{alicante},\ref{barcelona}}      \and 
    P.~Tanga\inst{\ref{oca}}               \and 
    F.~Vachier\inst{\ref{imcce}}           \and 
    A.~Vigan\inst{\ref{lam}}               \and 
    O.~Witasse\inst{\ref{estec}}           \and 
    B.~Yang\inst{\ref{eso}}                 
} 

   \institute{
     Institute of Astronomy, Faculty of Mathematics and Physics, Charles University, V~Hole{\v s}ovi{\v c}k{\'a}ch 2, 18000 Prague, Czech Republic
     \email{hanus.home@gmail.com, pepa@sirrah.troja.mff.cuni.cz}\label{prague}
     \and 
     Aix Marseille Univ, CNRS, LAM, Laboratoire d'Astrophysique de Marseille, Marseille, France
     \label{lam}
     \and
     Mathematics and Statistics, Tampere University, 33720 Tampere, Finland
     \label{tampere}
    \and 
     Space sciences, Technologies and Astrophysics Research Institute, Universit{\'e} de Li{\`e}ge, All{\'e}e du 6 Ao{\^u}t 17, 4000 Li{\`e}ge, Belgium
     \label{liege}
     \and 
     IMCCE, Observatoire de Paris, 77 avenue Denfert-Rochereau, F-75014 Paris Cedex, France
     \label{imcce}
     \and 
     Astronomical Observatory Institute, Faculty of Physics, Adam Mickiewicz University, ul. S{\l}oneczna 36, 60-286 Pozna{\'n}, Poland
     \label{poznan}
     \and 
     Universit\'e C{\^o}te d'Azur, Observatoire de la C{\^o}te d'Azur, CNRS, Laboratoire Lagrange, France
     \label{oca}
     \and 
     Department of Earth, Atmospheric and Planetary Sciences, MIT, 77 Massachusetts Avenue, Cam- bridge, MA 02139, USA
     \label{mit}
     \and 
     SETI Institute, Carl Sagan Center, 189 Bernado Avenue, Mountain View CA 94043, USA 
     \label{seti}
     \and 
     Center for Solar System Studies, 446 Sycamore Ave., Eaton, CO 80615, USA
     \label{warner}
     \and 
     Geneva Observatory, CH-1290 Sauverny, Switzerland
     \label{behrend}
     \and 
     Asociaci\'on Astron\'omica Astro Henares, Centro de Recursos Asociativos El Cerro C/ Manuel Aza{\~n}a, 28823 Coslada, Madrid, Spain
     \label{asenjo}
     \and 
     490 chemin du gonnet, F-38440 Saint Jean de Bournay, France
     \label{berger}
     \and 
     Institute of Geology, A. Mickiewicz University, Krygowskiego 12, 61-606 Pozna{\'n}
     \label{poznan2}
     \and 
     Observatoire de Durtal, F-49430 Durtal, France
     \label{charbonnel}
     \and 
     Observatorio Astron\'omico de C\'ordoba, C\'ordoba, Argentina
     \label{cordoba}
     \and 
     20 Parc des Pervenches, F-13012 Marseille, France
     \label{coliac}
     \and 
     Instituto de Astrofisica de Andalucia -- CSIC. Glorieta de la Astronom\'ia s/n, 18008 Granada, Spain
     \label{duffard}
     \and 
     I64, SL6 1XE Maidenhead, UK
     \label{jones}
     \and 
     Uranoscope, Avenue Carnot 7, F-77220 Gretz-Armainvilliers, France
     \label{leroy}
     \and 
     Anunaki Observatory, Calle de los Llanos, 28410 Manzanares el Real, Spain
     \label{molina}
     \and 
     Universidad de La Laguna, Dept. Astrofisica, E.38206 La Laguna, Tenerife, Spain
     \label{tenerife}
     \and 
     Institute of Theoretical Physics, Faculty of Mathematics and Physics, Charles University, V~Hole{\v s}ovi{\v c}k{\'a}ch 2, 18000 Prague, Czech Republic
     \label{teorka}
     \and 
     Courbes de rotation d'ast\' ero\" ides et de com\` etes, CdR
     \label{riemis}
     \and 
     Institute for Astronomy, University of Hawai'i, 2680 Woodlawn Drive, Honolulu, HI 96822, USA
     \label{hawaii}
     \and 
     Observatorio Amanecer de Arrakis, Alcal\'a de Guada\'ira, Sevilla, Spain
     \label{soldan}
     \and 
     Konkoly Observatory, Research Centre for Astronomy and Earth Sciences, Hungarian Academy of Sciences, Konkoly Thege 15-17, H-1121 Budapest, Hungary
     \label{konkoly}
     \and 
     Thirty-Meter-Telescope, 100 West Walnut St, Suite 300, Pasadena, CA 91124, USA
     \label{tmt}
     \and 
     Jet Propulsion Laboratory, California Institute of Technology, 4800 Oak Grove Drive, Pasadena, CA 91109, USA
     \label{jpl}
     \and 
     European Space Agency, ESTEC - Scientific Support Office, Keplerlaan 1, Noordwijk 2200 AG, The Netherlands
     \label{estec}
     \and 
     Open University, School of Physical Sciences, The Open University, MK7 6AA, UK
     \label{ou}
     \and 
     Laboratoire Atmosph\`eres, Milieux et Observations Spatiales, CNRS \& UVSQY, Guyancourt, France
    \label{lamos}
     %
     %
     \and 
     Departamento de F\'isica, Ingenier\'ia de Sistemas y Teor\'ia de la Se{\~n}al, Universidad de Alicante, E-03080 Alicante, Spain
     \label{alicante}
     \and 
     Institut de Ci\`encies del Cosmos, Universitat de Barcelona (IEEC-UB), Mart\'i i Franqu\`es 1, E-08028 Barcelona, Spain
     \label{barcelona}
     \and 
     European Southern Observatory (ESO), Alonso de Cordova 3107, 1900 Casilla Vitacura, Santiago, Chile
     \label{eso}
}

   \date{Received x-x-2016 / Accepted x-x-2016}
 
  \abstract
   {With an estimated diameter in the 320 to 350 km range, (704) Interamnia is the fifth largest main belt asteroid and one of the few bodies that fills the gap in size between the four largest bodies with $D>400$~km (Ceres, Vesta, Pallas and Hygiea) and the numerous smaller bodies with diameter $\leq$200 km. However, despite its large size, little is known about the shape and spin state of Interamnia and, therefore, about its bulk composition and past collisional evolution. }
   {We aimed to test at what size and mass the shape of a small body departs from a nearly ellipsoidal equilibrium shape (as observed in the case of the four largest asteroids) to an irregular shape as routinely observed in the case of smaller ($D\leq$200 km) bodies. } 
   {We observed Interamnia as part of our ESO VLT/SPHERE large program (ID: 199.C-0074) at thirteen different epochs. In addition, several new optical lightcurves were recorded. These data, along with stellar occultation data from the literature, were fed to the All-Data Asteroid Modeling (\adam{}) algorithm to reconstruct the 3D-shape model of Interamnia and to determine its spin state. }
   {Interamnia's volume-equivalent diameter of \Diam~km implies a bulk density of $\rho$=\Dens~\sid, which suggests that Interamnia -- like Ceres and Hygiea -- contains a high fraction of water ice, consistent with the paucity of apparent craters. Our observations reveal a shape that can be well approximated by an ellipsoid, and that is compatible with a fluid hydrostatic equilibrium at the 2\,$\sigma$ level.}
    {The rather regular shape of Interamnia implies that the size/mass limit, under which the shapes of minor bodies with a high amount of water ice in the subsurface become irregular, has to be searched among smaller ($D\leq$300km) less massive ($m\leq$3x10$^{19}$ kg) bodies. }

  \maketitle

\section{Introduction}\label{sec:introduction}

Because of their large masses, Solar-System bodies with diameters larger than $\sim$900~km possess rounded, ellipsoidal shapes, consistent with hydrostatic equilibrium. On the other side of the mass range, very small bodies (diameters $\leq$100~km) tend to possess highly irregular shapes, with the notable exception of some $D\leq$5~km bodies that are affected by the so-called YORP effect \citep[Yarkovsky–O'Keefe–Radzievskii–Paddack,][]{Rubincam2000, Vokrouhlicky2003}, and which have similar shapes to a spinning top \citep[e.g., Ryugu, or Bennu,][]{Watanabe2019,Nolan2013}.
The theory of the hydrostatic equilibrium of homogeneous bodies is well established \citep[e.g.,][]{Chandrasekhar1969}, whereas for differentiated bodies, approaches based on Clairaut equations \citep[e.g.,][]{Dermott1979, Chambat2010, Rambaux2015, Rambaux2017} or the numerical non-perturbative method are still under developement \citep{Hubbard2013}. From an observational point of view, it remains to be tested at what size range the shape of a typical minor body transits from a nearly rounded equilibrium shape to an irregular shape and to what extent this size range depends on factors such as the bulk composition of the minor planet or its collisional and thermal history.

Investigating these questions is one of the main motivations of our European Southern Observatory (ESO) large program (id: 199.C-0074; \citealt{Vernazza2018}) of which the aim is to constrain the shape of the forty largest main-belt asteroids. So far, our program has revealed that (10)~Hygiea, the fourth largest main-belt asteroid (D$\sim$434 km) possesses a shape that is nearly as spherical as that of (1)~Ceres \citep{Vernazza2019}, while being twice as small, whereas $D\sim$100--200~km bodies [(89)~Julia, (16)~Psyche, (41)~Daphne] possess irregular shapes \citep{Vernazza2018, Viikinkoski2018, Carry2019}. Asteroid (7)~Iris (D$\sim$214~km) is an intermediate case, as its shape appears to be consistent with that of an oblate spheroid with a large equatorial excavation \citep{Hanus2019a}.

Asteroid (704)~Interamnia, the fifth largest body in the main belt with a volume equivalent diameter in the 320--350 km size range \citep{Drummond2009a, Masiero2014}, is one of the very few asteroids that fills the gap in size between Hygiea and $D\sim$250~km-sized bodies. The remaining main belt asteroids in this size range are (31)~Euphrosyne \citep[$D$=282$\pm$10~km,][]{Masiero2013}, (52)~Europa \citep[$D$=314$\pm$5~km,][]{Hanus2017b}, (65)~Cybele \citep[$D$=296$\pm$25~km,][]{Viikinkoski2017}, and (511)~Davida \citep[$D$=311$\pm$5~km,][]{Viikinkoski2017}. Shapes of Europa and Davida already show some departures from a rotational triaxial ellipsoid \citep{Conrad2007, Merline2013}. Interamnia thus appears as another key target for investigating at what size the shape of a small body becomes irregular.

So far, little is known about Interamnia. It lacks a dynamical family, implying that it avoided a giant impact over the last $\sim$3 Gyrs. It was classified in the B-spectral class following the Bus taxonomy based on visible data alone, whereas it is labeled a C-type in the Bus-DeMeo taxonomy \citep{Clark2010b}. In the visible and near-infrared spectral range, it thus appears similar to objects such as (1)~Ceres, (10)~Hygiea, (24)~Themis, and (52)~Europa, which have been connected to interplanetary dust particles (IDPs) rather than to carbonaceous chondrites \citep{Vernazza2015, Vernazza2017, Marsset2016b}. Interamnia is also of great interest to the present study, being the largest asteroid for which a detailed shape model (convex or with local topography) and consistent spin-state solutions do not yet exist. This may be due to its shape being rather spherical, as suggested by the small brightness variations in its lightcurves \citep{Tempesti1975, Warner2018} and by the Keck disk-resolved images obtained by \citet{Drummond2008}.

Here, we present high-angular resolution imaging observations of Interamnia with ESO VLT/SPHERE/ZIMPOL that were performed as part of our large program. We use these observations along with several newly acquired lightcurves to constrain its 3D shape and its spin for the first time.

\section{Observations}\label{sec:data}

The observations used in our analysis of physical properties of Interamnia consist of disk-resolved images from the VLT telescope and disk-integrated optical lightcurves from various sources, including our observing campaign.

\subsection{Disk-resolved data}\label{sec:ao}

Interamnia was observed with VLT/SPHERE/ZIMPOL \citep[Spectro-Polarimetric High-contrast Exoplanet REsearch, Zurich IMaging POLarimeter,][]{Thalmann2008} in the narrow band imaging configuration (N$\_$R filter; filter central wavelength = 645.9 nm, width = 56.7 nm) during two consecutive apparitions in August--September 2017, and between December 2018 and January 2019. During both apparitions, the angular size of Interamnia was in the 0.20--0.26'' range, with a slightly larger angular size during the first apparition. Interamnia's extent on the images reaches up to 80~pixels. Both datasets sample the whole rotation phase of Interamnia, although not as evenly as initially expected (we recall here that our nominal observing strategy is to image our large program targets every 60 degrees throughout their rotation). Nonetheless, the satisfactory rotation phase coverage, along with a nearly equator-on geometry during both apparitions lead to a nearly complete surface coverage ($\sim$95\%) that makes it possible to constrain the three dimensions of Interamnia well. 

The reduced images were deconvolved with the \mistral~algorithm \citep[see][for details about the deconvolution procedure]{Fetick2019} and are shown in Figs.~\ref{fig:Deconv1} and~\ref{fig:Deconv2}. Table~\ref{tab:ao} contains full information about the data.

Finally, in addition to the AO data, we also utilized four stellar occultations obtained in 1996, 2003, 2007, and 2012. However, the stellar occultations are largely redundant given the coverage of our SPHERE observations. We provide details about these observations in Table~\ref{tab:occ}.

\subsection{Optical photometry}\label{sec:photometry}

\setkeys{Gin}{draft=false}
\begin{figure}
\begin{center}
\resizebox{1.0\hsize}{!}{\includegraphics{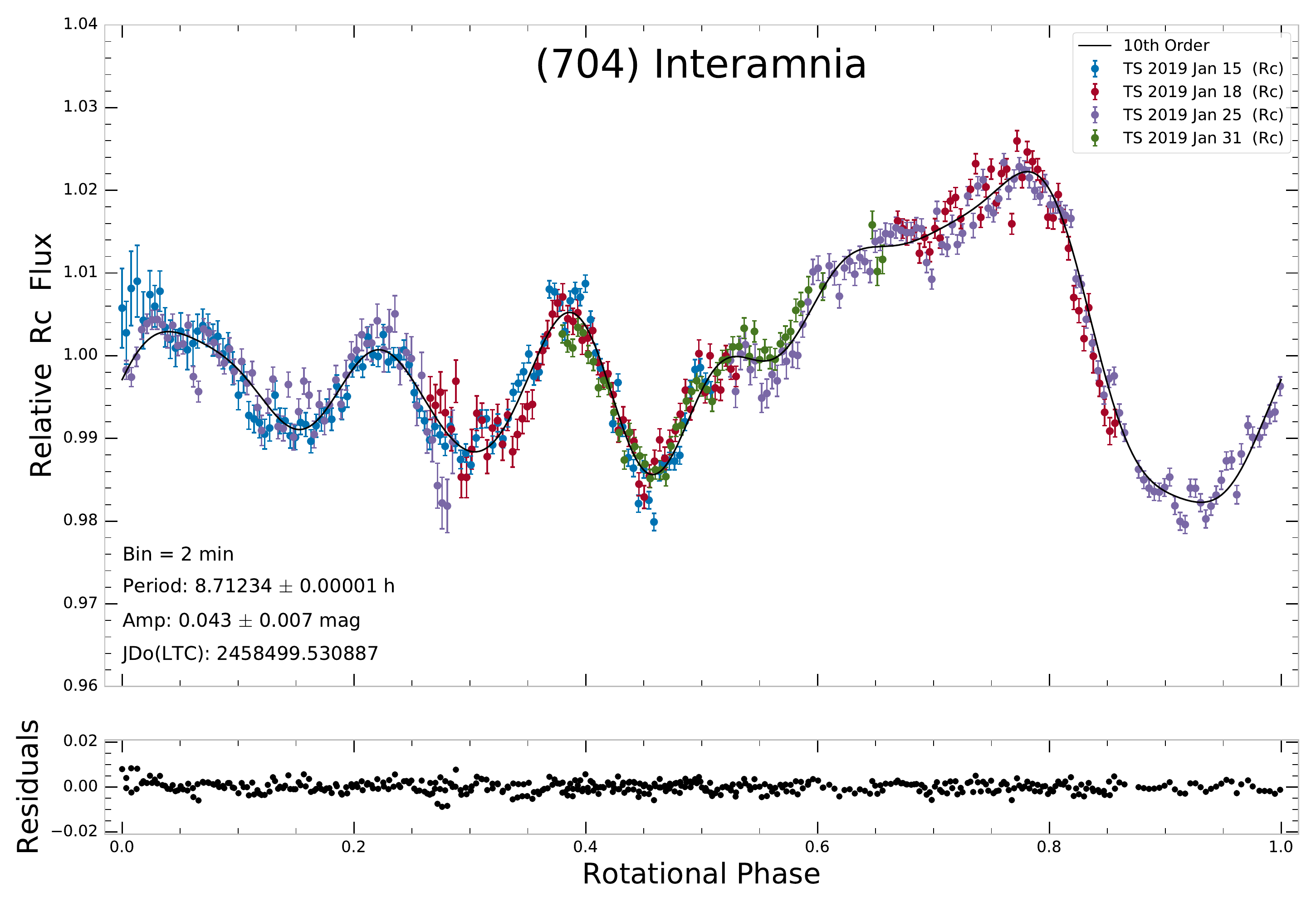}}\\
\end{center}
\caption{\label{fig:trappist}Upper panel: Composite lightcurve of (704)~Interamnia obtained with TRAPPIST-South telescope. A Fourier series of tenth order is fitted to the data. Lower panel: Residuals of the fit.}
\end{figure}
\setkeys{Gin}{draft=true}

We compiled a large dataset of 189 optical lightcurves sampling 15 apparitions. These data include lightcurves downloaded from the Asteroid Photometric Catalog \citep[APC,][]{Piironen2001} with original references: \citet{Tempesti1975, Lustig1976, Shevchenko1992, Michalowski1995b}. Many lightcurves were also provided through the courbes de rotation d'ast\' ero\" ides et de com\` etes database (CdR\footnote{\texttt{http://obswww.unige.ch/\textasciitilde behrend/page\_cou.html}}), maintained by Raoul Behrend at the  Observatoire de Gen\` eve, and through the ALCDEF\footnote{\url{http://http://alcdef.org/}} database maintained by Brian Warner \citep{Warner2018}. The largest photometric dataset was obtained from the SuperWASP archive \citep{Grice2017}: 114 lightcurves spanning years 2006--2011. Finally, one lightcurve was obtained at Wallace Observatory, and a densely covered dataset was obtained by the TRAPPIST-South and -North \citep{Jehin2011} as a support for this study.

Additional photometric data were gathered within the observing campaign of the "Small Bodies: Near And Far" project \citep{Muller2018}, with partial participation of Gaia-GOSA observers. Gaia-GOSA\footnote{\url{www.gaiagosa.eu}} is a web service dedicated to amateur observers willing to support asteroid studies through targeted photometric campaigns. The website makes it possible to coordinate a worldwide observing campaign, which is especially important for slow rotating objects requiring long-term observations over several nights. Interamnia was observed on Gaia-GOSA during its last two apparitions (2017 and 2018), providing new lightcurves for our dataset. 

Finally, we also made use of sparsely sampled $V$-band photometry from the All-Sky Automated Survey for Supernovae \citep[ASAS-SN,][]{Shappee14,Kochanek2017} and Gaia Data Release 2 \citep[DR2,][]{Spoto2018}. The ASAS-SN data sample five consequent apparitions between 2013 and 2018, and contain 196 individually calibrated measurements in the Johnson V band. Gaia DR2 data are internally calibrated, however, they are also limited to only 16 individual measurements. The sparse data are processed following the same procedures applied, for example, in \citet{Hanus2011}, or \citet{Durech2018d}. Other sparsely sampled data used so far for the shape modeling \citep[e.g., USNO-Flagstaff, Catalina Sky Survey, Lowell,][]{Durech2009, Hanus2011, Hanus2013a, Durech2016} have photometric uncertainties, at best, comparable to the lightcurve amplitude of Interamnia (usually $<$0.1 mag), which makes them useless for the shape modeling. On the other hand, the high-photometric precision of the ASAS-SN data ($\sim$0.04 mag) and Gaia DR2 ($\sim$0.02 mag) implies that brightness changes due to irregular shape and spin state are distinguishable from the photometric noise. Sparse data are particularly useful for the spin-state determination, because they cover a large range of observing geometries (i.e., phase angles).

The basic characteristics of the photometric data are listed in Table~\ref{tab:lcs}. In general, Interamnia's lightcurves exhibit a brightness variation pattern consistent with a synodic rotation period of $\sim$8.7 h and rather small amplitude of the brightness changes within the rotation (usually $<$0.1 mag, see, e.g., Fig.~\ref{fig:trappist}). These small changes make the determination of a unique shape model and spin-state solution challenging.

\section{Results\\ }\label{sec:results}

The rich datasets of disk-integrated optical lightcurves and VLT/SPHERE/ZIMPOL disk-resolved images enabled us to derive the convex shape model of Interamnia, as well as its 3D-shape model with local topography. Moreover, we also estimated Interamnia's bulk density and analysed its shape with respect to the hydrostatic equilibrium. Finally, we discuss few identified surface features.

\subsection{Spin-state determination by convex inversion}\label{sec:convexinv}

\begin{table*}
 \caption{\label{tab:spins}Summary of published spin-state solutions for Interamnia. The table gives the ecliptic longitude $\lambda$ and latitude $\beta$ of all possible pole solutions with their uncertainties, the sidereal rotation period $P$, and the reference. Our second pole solution from convex inversion (CI) has been rejected due to inconsistency with the SPHERE images.}
 \centering
 \begin{threeparttable}
\begin{tabular}{cccc c l} \hline
\multicolumn{1}{c} {$\lambda_1$} & \multicolumn{1}{c} {$\beta_1$} & \multicolumn{1}{c} {$\lambda_2$} & \multicolumn{1}{c} {$\beta_2$} & \multicolumn{1}{c} {$P$} & Note \\
\multicolumn{1}{c} {[deg]} & \multicolumn{1}{c} {[deg]} & \multicolumn{1}{c} {[deg]} & \multicolumn{1}{c} {[deg]} & \multicolumn{1}{c} {[hours]} &  \\
\hline\hline
 $43\pm8$  & $-21\pm9$  & $224\pm10$  & $-22\pm10$ & --                   & \citet{Michalowski1993} \\
 $47\pm10$ & $-3\pm10$  & $227\pm10$  & $1\pm10$   & --                   & \citet{DeAngelis1995b} \\
 $51\pm15$ & $22\pm10$  &             &            & $\phantom{4}8.72729\pm0.00001\phantom{4}$ & \citet{Michalowski1995b} \\
 $36\pm5$  & $12\pm5$   &             &            & --                   & \citet{Drummond2008} \\
 $47\pm10$ & $66\pm10$  &             &            & --                   & \citet{Drummond2009a} \\
 $259\pm8$ & $-50\pm5$  &             &            & 8.728967167$\pm$0.000000007$^a$       & \citet{Sato2014} \\
 $87\pm10$ & $63\pm10$  & \sout{$226\pm10$}  & \sout{$43\pm10$} & $\phantom{3}8.712355\pm0.000005\phantom{3}$  & This work, CI \\
 $87\pm5$  &  $62\pm5$  &             &            & $\phantom{4}8.71234\pm0.00001\phantom{4}$  & This work, \adam{} \\
\hline
\end{tabular}
 \begin{tablenotes}[para,flushleft]
     \centering $^a$ Such uncertainty in the rotation period is unrealistic. 
 \end{tablenotes}
 \end{threeparttable}
\end{table*}

To derive the first reliable shape and spin-state solution for Interamnia, we implemented the standard convex inversion method of \citet{Kaasalainen2001a} and \citet{Kaasalainen2001b} that takes disk-integrated data as the only data input, and searches the set of parameters describing the shape and rotation state that best match the data. The search was done on a grid of parameters, where each set of input parameters converges to a local minimum in the parameter space. We tested all reasonable combinations of relevant parameters to find those that correspond to the global minimum. In convex inversion, the shape is parametrized by a convex polyhedron, the rotation state is described by the sidereal rotation period, and the ecliptic coordinates (longitude and latitude) of the spin axis, and we used a simple three-parameter phase function relation that is necessary when sparse data are included \citep{Kaasalainen2001b}. The convex inversion procedure essentially consists of two parts. Firstly, we ran the convex inversion for rotation periods from the 8.6--8.8~h interval, which contains all previous estimates that concentrate near $\sim$8.72~h. The step in the rotation period was selected in a way that each local minimum is sampled \citep{Kaasalainen2001b}. For each period, we ran the convex inversion with 10 different initial pole orientations isotropically distributed on a sphere and selected only the best fitting solution. Then, we constructed the dependence of the rms value on the sampled period in Fig.~\ref{fig:period}. This periodogram has a clear minimum near 8.71~h, moreover, only one period value ($P$=8.71236~h) provides a significantly better fit to the observed data than all the other periods. We applied the same criteria as in \citet[][ for more details and additional references]{Hanus2018c} to distinguish between acceptable solutions and those that should already be rejected. Secondly, we ran the convex inversion with the unique period found in the previous step, with a higher shape model resolution, and for many pole orientations ($\sim$50) isotropically distributed on a sphere. Only four pole solutions fell within the rms limit from \citet{Hanus2018c}. Moreover, two solutions out of the four with the worst fit had non-physical shapes with their maximum moment of inertia significantly nonaligned with the rotation axis. Therefore, we derived only two possible spin-state and shape solutions, which we list in Table~\ref{tab:spins}. 

Our spin-state solutions are rather different from those previously published with the main disagreement in the ecliptic latitude -- we found a prograde rotation with ecliptic latitude of $\sim$40--60$^\circ$, while previous authors derived mostly smaller values between --20$^\circ$ and 20$^\circ$ \citep{Michalowski1993, Michalowski1995b, DeAngelis1995b, Drummond2008}, or even --50$^\circ$ \citep{Sato2014} for the latitude. On the other hand, the determinations for the ecliptic longitude are mostly consistent with each other. The closest solution to ours is from \citet{Drummond2009a} based on disk-resolved images from Keck. The rather significant differences are likely caused by (i)~Interamnia having small brightness variations, and (ii)~the fact that the spin-state determination based on photometric data with low S/N ratio is challenging and could lead to inaccurate determinations \citep[or to underestimated uncertainties,][]{Marciniak2015}.
Our first reliable spin-state solution of Interamnia is used as an input for the shape modeling with \adam{} in the following section.

\subsection{3D-shape reconstruction with \adam{}}\label{sec:ADAM}

\setkeys{Gin}{draft=false}
\begin{figure*}
\begin{center}
\resizebox{0.999\hsize}{!}{\includegraphics{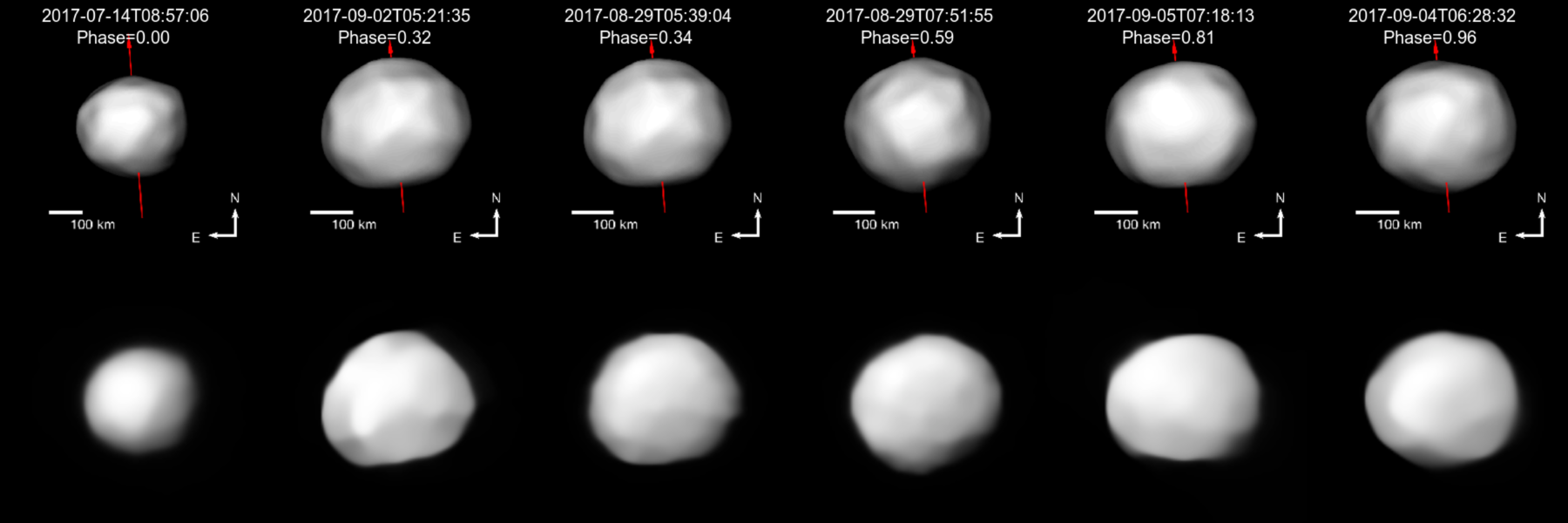}}\\
\vspace{-1mm}
\resizebox{0.999\hsize}{!}{\includegraphics{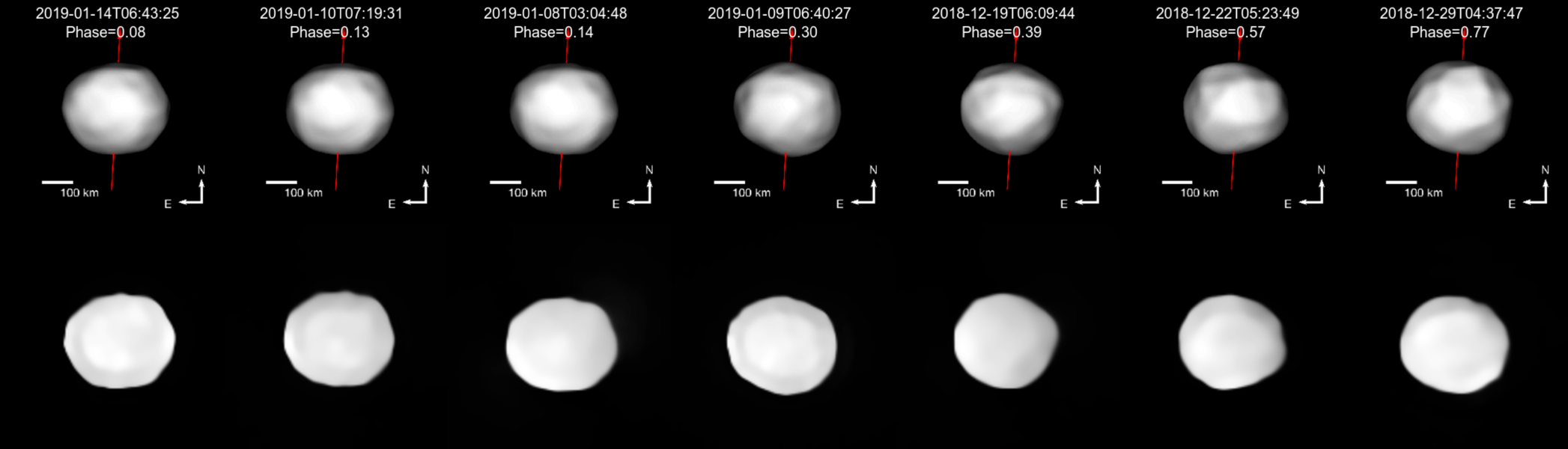}}\\
\end{center}
\caption{\label{fig:comparison}Comparison between VLT/SPHERE/ZIMPOL deconvolved images of Interamnia (second and fourth rows) and the corresponding projections of our \adam{} shape model (first and third row). The red line indicates the position of the rotation axis. We use a nonrealistic illumination to highlight the local topography of the model.}
\end{figure*}
\setkeys{Gin}{draft=true}

\setkeys{Gin}{draft=false}
\begin{figure}
\begin{center}
\includegraphics[width=0.24\textwidth]{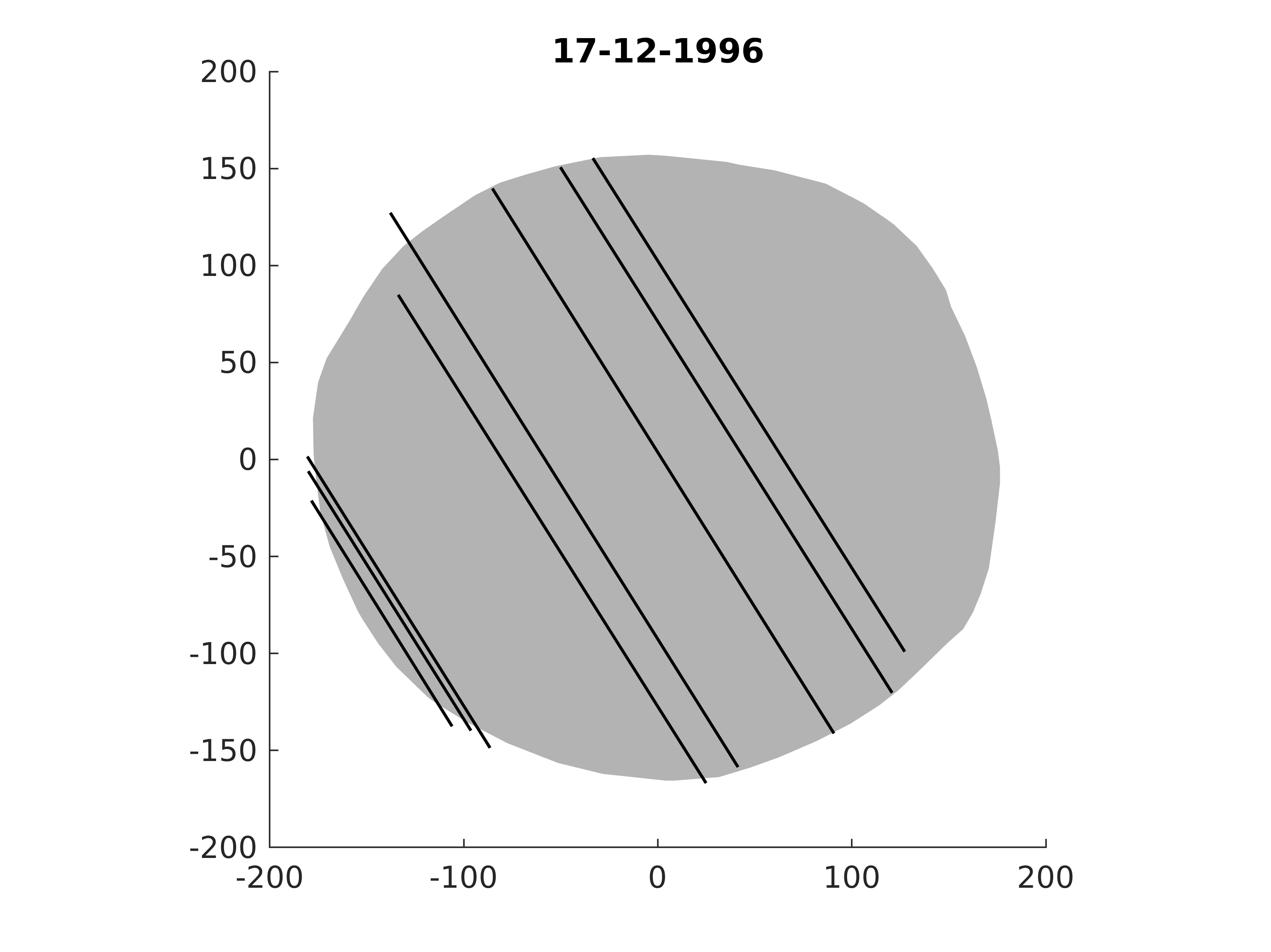}\includegraphics[width=0.24\textwidth]{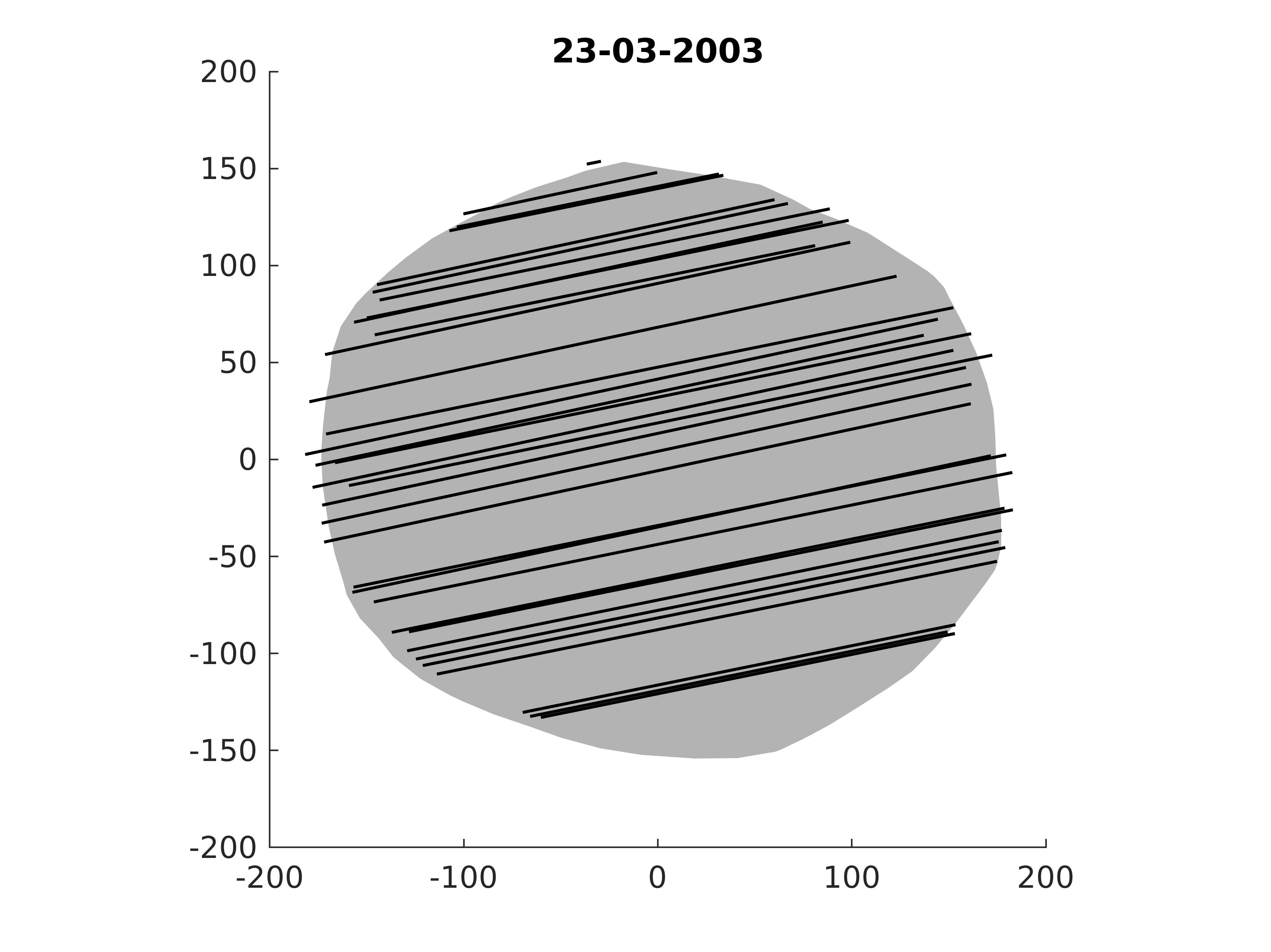}\\
\includegraphics[width=0.24\textwidth]{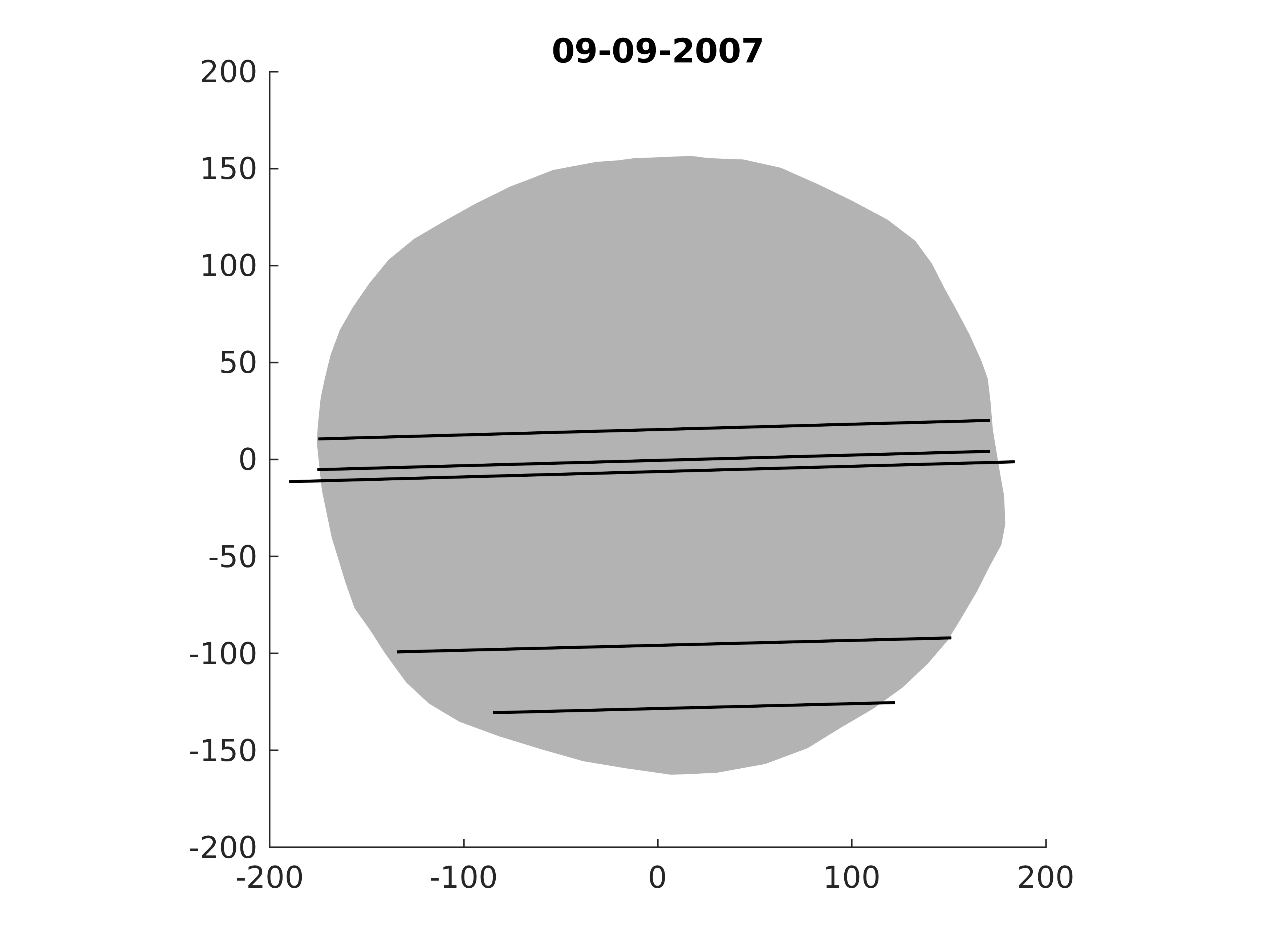}\includegraphics[width=0.24\textwidth]{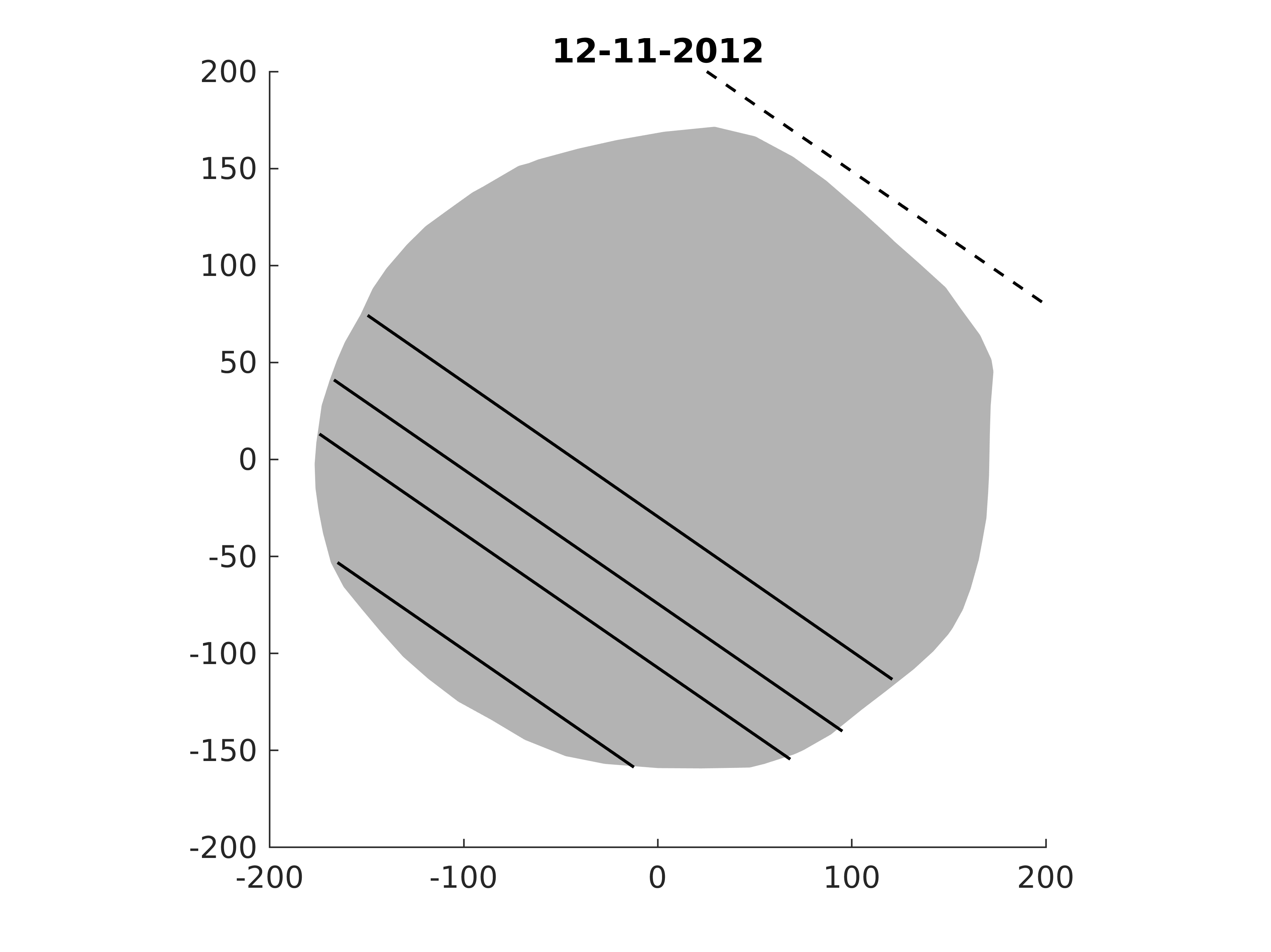}\\
\end{center}
\caption{\label{fig:occ}Observed occultation chords and model silhouettes. The dashed line is a negative observation. North is up and East to the left. The axis scale corresponds to kilometers.}
\end{figure}
\setkeys{Gin}{draft=true}

The disk-resolved data allow us to perform the 3D-shape optimization with the \adam{} algorithm. Firstly, we compared the SPHERE images with corresponding projections of the two shape solutions derived by the convex inversion and found that only one solution is consistent with the images (see Table.~\ref{tab:spins}), therefore, we continued the shape modeling only with the preferred rotation state solution as an input.

We proceeded with the modeling the standard way \citep[see, e.g.,][]{Viikinkoski2018, Hanus2019a}: we constructed a low-resolution shape model using the octantoid \citep{Viikinkoski2015} shape parametrization while balancing the fit to the optical lightcurves, SPHERE images, and stellar occultations. We applied the \adam{} algorithm to a dataset of 189 optical lightcurves, 60 VLT/SPHERE/ZIMPOL images, and four stellar occultations. Then, we increased the shape model resolution and the weight of the SPHERE data with respect to the lightcurves and occultations, and used the low-resolution shape model as an initial input. We tested several combinations of shape resolutions and relative weighting of the observed data to confirm the consistency of our results. The comparison between the shape model projections and SPHERE observations is shown in Fig.~\ref{fig:comparison}. Moreover, we also provide the model fit to the stellar occultations in Fig.~\ref{fig:occ}. All four stellar occultations agree well with our shape model.

The physical properties of our final solution are listed in Table~\ref{tab:param}. The uncertainties reflect the typical ranges of parameters within the various individual solutions we obtained (for different shape resolutions, relative data weights). The uncertainties are also consistent with the size of one to two pixels. The volume-equivalent diameter of Interamnia  (\Diam~km) is well constrained because of the equator-on observations during both apparitions, and because the overall coverage of the AO observations amounts to $\sim95$\% of the model surface area. Moreover, for the same reason, the $c$ dimension is also reliably estimated, which happens rather rarely, because the usually limited geometry coverage of the images makes the determination of the $c$ dimension inaccurate. Our size estimate is larger than those of \citet{Drummond2009a} and \citet{Sato2014}, but both are in agreement with ours within the 1\,$\sigma$ uncertainties. 

The shape model along with the lightcurve data and the fit to all datasets have been uploaded to the online Database of Asteroid Models from Inversion Techniques \citep[DAMIT\footnote{\url{http://astro.troja.mff.cuni.cz/projects/damit}},][]{Durech2010}.

\subsection{Density}\label{sec:Density}

We combined the derived volume of Interamnia with the best estimate of its mass \Mass~kg (Table~\ref{tab:mass} and Fig.~\ref{fig:mass}) and obtained a bulk density estimate of \Dens~\sid. Our mass estimate is based on all relevant determinations collected in the literature \citep[as we did in our previous studies, for instance,][]{Carry2012,Vernazza2018, Viikinkoski2018, Hanus2019a}.

The relative uncertainty affecting the bulk density of Interamnia exceeds 30\%, preventing us from drawing meaningful conclusions about the body's composition. For instance, the bulk density is compatible within 1\,$\sigma$ error with those of the two largest C-type asteroids, Ceres \citep[2.161$\pm$0.003 \sid,][]{Park2019} and Hygiea \citep[1.94$\pm$0.25 \sid,][]{Vernazza2019}, but also of silicate bodies such as (25143)~Itokawa \citep[1.90$\pm$0.13 \sid,][]{Fujiwara2006} or (433)~Eros \citep[2.67$\pm$0.10 \sid,][]{Veverka2000}. Current estimates of the densities of asteroids with masses greater than $\sim$5\,$\times$\,$10^{18}$ kg imply a small amount of macroporosity within these bodies \citep{Carry2012b, Viikinkoski2015b, Marsset2017, Carry2019, Hanus2019a}, and spectroscopic observations of Interamnia in the 3-micron region have revealed the presence of hydrated material at its surface \citep{Usui2019} and spectral similarity to Ceres \citep{Rivkin2019}. Therefore, we can assume that Interamnia's bulk density is close to that of Ceres. This implies a large amount of water inside Interamnia, likely as a mixture of ice and phyllosilicates, as in the case of Ceres.

\begin{table*}
 \caption{\label{tab:param}
 Volume-equivalent diameter ($D$), dimensions along the major axis ($a$, $b$, $c$), sidereal rotation period ($P$), spin-axis ecliptic J2000 coordinates (longitude $\lambda$ and latitude $\beta$), mass ($m$), and bulk density ($\rho$) of Interamnia as determined here, compared with the work of \citet{Drummond2008, Drummond2009a, Sato2014}. Uncertainties correspond to 1\,$\sigma$ values.}
 \centering
 \begin{threeparttable}
 \begin{tabular}{llcccc}
  \hline
  Parameter  & Unit & \citet{Drummond2008} & \citet{Drummond2009a} & \citet{Sato2014} & This work, \adam{}\\
  \hline
   $D$       & km   &                & 319$\pm$9         & 327$\pm$3         & \Diam  \\
   $\lambda$ & deg. & 36$\pm$6       & 47$\pm$3          & 259$\pm$8         & 87$\pm$5  \\
   $\beta$   & deg. & 12$\pm$11      & 66$\pm$3          & --50$\pm$5        & 62$\pm$5   \\
   $P$       & h    &                &  8.727            & 8.728967167(7)    & 8.712336(10) \\
   $a$       & km   & 385$\pm$53     & 349$\pm$4         & 362$\pm$3         & 362$\pm$8 \\
   $b$       & km   & 337$\pm$21     & 339$\pm$3         & 324$\pm$5         & 348$\pm$8  \\
   $c$       & km   & 163$\pm$184    & 274$\pm$22        & 297$\pm$4         & 310$\pm$8  \\
   $a/b$     &      & 1.14$\pm$0.17  & 1.03$\pm$0.01     & 1.15$\pm$0.02     & 1.04$\pm$0.02  \\
   $b/c$     &      & 2.1$\pm$2.3    & 1.24$\pm$0.10     & 1.20$\pm$0.01     & 1.13$\pm$0.02 \\
   $m$       & $10^{19}$~kg &        &                   & 6.96$\pm$1.79$^a$ & \Masss  \\
   $\rho$    & \sid &                &                   & 3.8$\pm$1.0       & \Dens \\
  \hline
 \end{tabular}
 \begin{tablenotes}[para,flushleft]
     \centering $^a$ \citet{Michalak2001}. 
 \end{tablenotes}
 \end{threeparttable}

\end{table*}

\subsection{Shape analysis}\label{sec:shape1}

\setkeys{Gin}{draft=false}
\begin{figure}
\begin{center}
\resizebox{0.90\hsize}{!}{\includegraphics{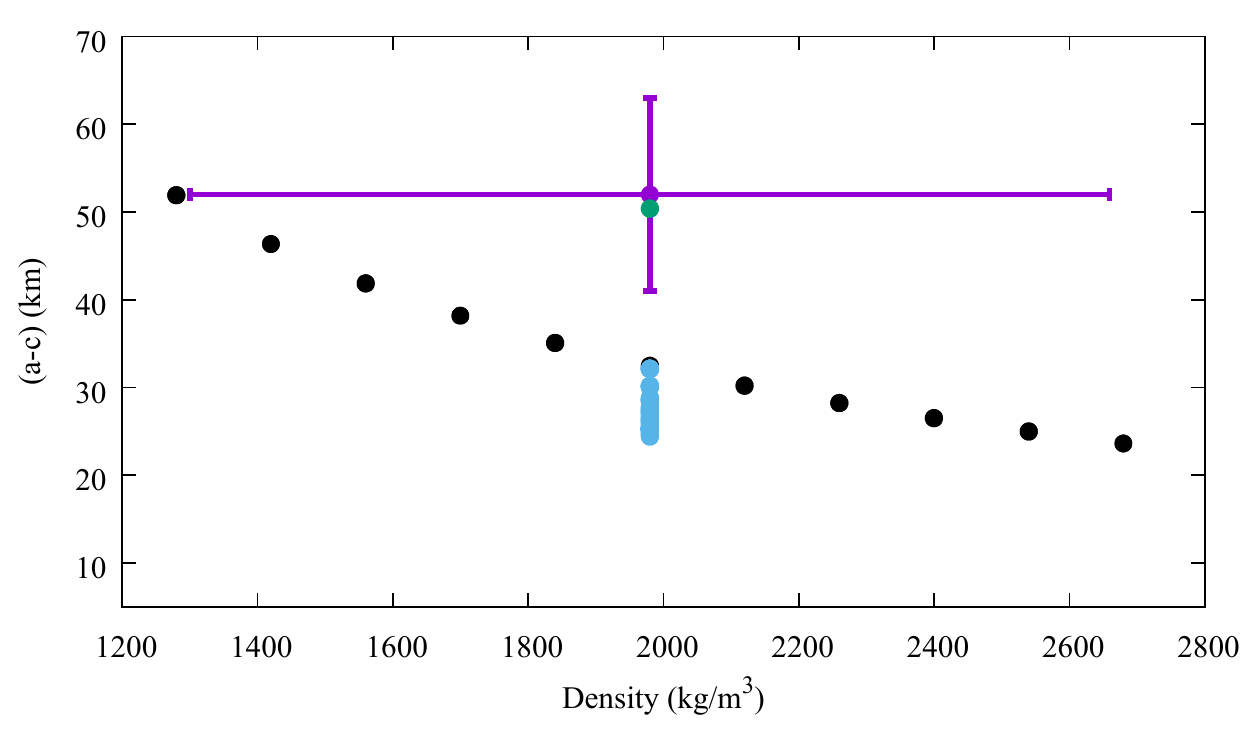}}\\
\end{center}
\caption{\label{fig:equilibrium}Results of ($a-c$) for Interamnia as a function of mean density for homogeneous case (sequence of black dots) and core-mantle differentiated case (blue vertical dots). For the latter case, we explored the 1\,100--1\,600~\sidd \,range for the mantle density, and the 2\,050--3\,400~\sidd \,range for the core density while keeping the bulk density at a constant value of 1\,980~\sidd. The value of $a-c$ decreases when mass increases with depth. The purple cross represents the value derived in Sect.~\ref{sec:shape1} with its 1\,$\sigma$ uncertainty (uncertainties of $a$ and $b$ are added quadratically). Finally, the green dot represents the values of ($a-c$) for the homogeneous case and Interamnia's assumed faster rotation period of 7.1~h.}
\end{figure}
\setkeys{Gin}{draft=true}

As a first step, we performed an analysis of Interamnia's shape, similar to the one performed in the case of Hygiea \citep{Vernazza2019}. We fitted Interamnia’s 3D-shape model with an ellipsoid and subsequently measured the radial difference between the two shapes. It appears that the large-scale topography of Interamnia is very subdued, without noticeable large impact basins on its surface (Fig.~\ref{fig:histogram}), similar to that of Ceres and Hygiea \citep{Vernazza2019}. As in the case of Hygiea, the relative volume difference between Interamnia's 3D-shape model and that of its best-fitting ellipsoid is 0.2\%, which implies that Interamnia's shape is very close to that of an ellipsoid. Next, we calculated the sphericity of Interamnia as done previously in the case of Hygiea \citep{Vernazza2019}. We found a sphericity value of 0.9880, similar to that of Vesta (0.9860), and only marginally lower than that of the nearly spherical dwarf planet candidate Hygiea \citep[0.9975,][]{Vernazza2019}. 

Given the ellipsoidal shape of Interamnia and the fact that its $a$ and $b$ axes have similar lengths (within errors) and that the $c$ dimension is shorter than the $a$ and $b$ axes, we investigated whether Interamnia's shape may be at hydrostatic equilibrium. We investigated both (i) a homogeneous and (ii) a core-mantle differentiated case. Indeed, given the large uncertainty of the bulk density coming from the large mass uncertainty, both models are viable possibilities.

The hydrostatic equilibrium figure of an homogeneous body can be computed using MacLaurin’s equation \citep[e.g.,][]{Chandrasekhar1969}:
\begin{equation}
\frac{\Omega^2}{\pi G \rho} =  \frac{2 \sqrt{1-e^2}}{e^3} (3-2 e^2) \arcsin{(e)} -6 \frac{1- e^2}{e^2},
\label{eq:maclaurin}
\end{equation} 
where $G$ is the gravitational constant, $\Omega$ is the rotational velocity, and $e$ is the ellipticity of the body shape defined by
\begin{equation}
e^2 = 1 - \left(\frac{c}{a}\right)^2.
\end{equation}
The MacLaurin equation is not valid for a differentiated body \citep{Moritz1990}. In this case, the hydrostatic equilibrium figure can be derived through Clairaut’s equations developed to an order that depends on a parameter $m$ called geodetic parameter \citep{Chambat2010, Rambaux2015}:
\begin{equation}
m = \frac{\Omega^2 R^3}{G M},
\end{equation} 
where $\Omega$ is the angular spin velocity, $R$ the mean radius, and $M$ the mass of the body. Considering the particular value of $m$ and the quality of available observations, Clairaut’s equations may be developed to first, second, or third order \citep{Lanzano1974}. For example, at first order, Clairaut's equation is written as \citep{Kopal1960, Lanzano1974}
\begin{equation}
r^2 \ddot{f_2} + 6 \gamma r \dot{f_2} + 6 (\gamma-1) f_2  = 0,
\end{equation}
where $f_2$ corresponds to the coefficient of the Legendre polynomial of degree two of the equipotential surface~$s$
\begin{equation}
s(r,\theta) = r( 1+ f_2(r) P_2(\cos{\theta}) ),
\end{equation}
and $\gamma = \rho(r) / \bar{\rho(r)}$ is the ratio between the density of the layer at $r$ and the mean density at $r$. \citet{Lanzano1974} developed the equations up to order three by introducing the following coefficients:
\begin{equation}
s(r,\theta) = r( 1+ f_2(r) P_2(\cos{\theta}) + f_4(r) P_4(\cos{\theta}) + f_6(r) P_6(\cos{\theta})),
\end{equation}
and he obtained a system of three differential equations and boundary conditions (see the equations in \citealt{Lanzano1974}). A numerical scheme to solve these equations has previously been applied to the hydrostatic figures of Earth \citep{Chambat2010}, Ceres \citep{Rambaux2015,Park2016} and Pallas \citep{Marsset2019}, and now to a differentiated Interamnia.

For the homogeneous case, we computed $a-c$ values for mean densities within the 1\,300--2\,700~\sidd \,range, while for the core-mantle differentiated model, we explored the 1\,100--1\,600~\sidd \,range for the mantle density and the 2\,050--3\,400~\sidd \,range for the core density while keeping the bulk density at 1\,980~\sidd. As expected, $a-c$ decreases when mass increases with depth. In Fig.~\ref{fig:equilibrium}, we present the $a-c$ values as a function of mean density for both the homogeneous case (black dots) and the core-mantle differentiated case (blue dots) and compare them with the observed values. 

Assuming Interamnia's bulk density of 2\,000~\sidd\, (i.e., similar to that of Ceres or Hygiea), its shape is consistent with hydrostatic equilibrium at the 2\,$\sigma$ level. We further calculated, by aiming for the central value of $a-c$, that Interamnia's shape would be at hydrostatic equilibrium at the 1\,$\sigma$ level for a slightly shorter rotation period (7.1 h, see Fig.~\ref{fig:equilibrium}), assuming homogeneous interior. The core-mantle differentiated case requires slightly larger despinning. This is in agreement with the collisional models that predict statistical preference of despinning by impacts \citep{Sevecek2019}.

Overall, these results are compatible with a formation of Interamnia at hydrostatic equilibrium. Interamnia's global equilibrium shape is likely a consequence of both its large mass and the initial presence of a large amount of water ice in its interior. During its early history, a large fraction of the water ice would have melted due to the radioactive decay of $^{26}$Al implying the presence of liquid water in its interior, and thus an early fluid interior as is the case for Vesta, Ceres, and Hygiea \citep{Takir2012,Vernazza2017}.

\subsection{Surface topography}\label{sec:shape}

\begin{figure}
\begin{center}
\resizebox{0.90\hsize}{!}{\includegraphics{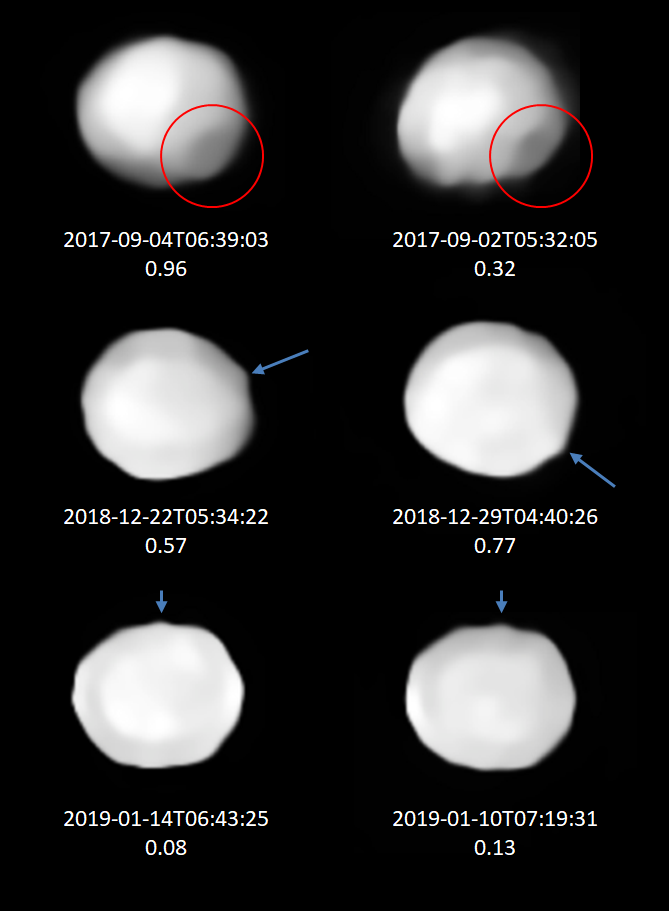}}\\
\end{center}
\caption{\label{fig:topography}Topographic features identified on Interamnia. The arrows indicate potential surface features (central peaks of large impact basins) and the red circles the two darker circular regions.}
\end{figure}

We observe only two large depressions (apparent dark regions) in the bottom-right parts of the images with rotation phases 0.32 and 0.96 (first apparition). In addition, a few mountain-like features can be observed in the object's contours. The most prominent one lies very close to the north pole and is visible at three epochs from the second apparition (rotation phases 0.08, 0.13 and 0.14). This feature could be a central peak of a $\sim$150--200 km large crater. Two similar topographic features are located to the bottom right of the image at rotation phase 0.77 (second apparition), and on the right of rotation phase 0.57 (Fig.~\ref{fig:topography}).

\setkeys{Gin}{draft=false}

Compared with the large topographic variations found on S-type asteroids such as (3)~Juno, (6)~Hebe, and (7)~Iris \citep{Viikinkoski2015b, Marsset2017, Hanus2019a}, Interamnia's surface appears relatively smooth with only a few basins or depressions. From this point of view, Interamnia appears very similar to Hygiea and Ceres \citep{Vernazza2019}. A plausible explanation for the lack of obvious craters at the resolution of these SPHERE images may be, as proposed in the case of Hygiea \citep{Vernazza2019}, that the craters are mostly complex flat-floored rather than simple bowl-shaped. The expected simple-to-complex crater transition diameter for Interamnia, assuming a water-rich composition for its mantle in agreement with our density estimate, should be around 30~km \citep{Hiesinger2016} (the transition diameter for a rock-dominated composition would be around 70~km). Given the spatial resolution of our observations (D$\sim$30--40 km), the paucity of large bowl-shaped craters on Interamnia can be attributed to its water-rich mantle composition. 

\section{Summary}\label{sec:conclusions}

We derived the first reliable spin-state solution of Interamnia. This success was only possible due to a large participation in our photometric campaign and data mining from survey telescopes (SuperWASP, ASAS-SN, Gaia). The role of observers with small aperture telescopes was essential.

Our 3D-shape model of Interamnia derived by \adam{} from the spectacular SPHERE disk-resolved images is nearly ellipsoidal with almost equal equatorial dimensions ($a/b$=1.04) and is only slightly flattened with $b/c$=1.13. Interamnia's shape appears to be in hydrostatic equilibrium at the 2\,$\sigma$ level. It follows that the size and mass limit under which minor bodies' shapes become irregular has to be searched among smaller (D$\leq$300~km) less massive ($m\leq$3x10$^{19}$ kg) bodies.

Our volume equivalent diameter of \Diam~km makes Interamnia the fifth largest object in the main belt after (1)~Ceres, (2)~Pallas, (4)~Vesta, and (10)~Hygiea. The other two 300-km-class bodies -- (52)~Europa and (65)~Cybele -- are likely smaller than Interamnia, although their size estimates have rather large uncertainties. Finally, spectroscopic observations in the near infrared and the bulk density of $\rho$=\Dens~\sid\, suggests that Interamnia -- like Ceres and Hygiea -- contains a high fraction of water in the subsurface. This would provide a natural explanation for the lack of obvious craters wider than a few tens of km, as well as for its ellipsoidal/regular shape, similarly to what has been proposed for Hygiea by \citet{Vernazza2019}.

\begin{acknowledgements}
This work has been supported by the Czech Science Foundation through grant 18-09470S (J.~Hanu\v s, J.~\v Durech) and by the Charles University Research program No. UNCE/SCI/023. This research was supported by INTER-EXCELLENCE grant LTAUSA18093 from the Czech Ministry of Education, Youth, and Sports (J.~Hanu\v s and O.~Pejcha). The research of O.~Pejcha is additionally supported by Horizon 2020 ERC Starting Grant ``Cat-In-hAT'' (grant agreement \#803158) and award PRIMUS/SCI/17 from Charles University. P.~Vernazza, A.~Drouard, and B.~Carry were supported by CNRS/INSU/PNP. M. Marsset was supported by the National Aeronautics and Space Administration under Grant No. 80NSSC18K0849 issued through the Planetary Astronomy Program. This work was supported by the National Science Centre, Poland, through grant no. 2014/13/D/ST9/01818 (A.~Marciniak). The research leading to these results has received funding from the European Union's Horizon 2020 Research and Innovation Programme, under Grant Agreement no 687378 (SBNAF). This project has been supported by the GINOP-2.3.2-15-2016-00003 and NKFIH K125015 grants of the Hungarian National Research, Development and Innovation Office (NKFIH) and by the Lend\"ulet grant LP2012-31 of the Hungarian Academy of Sciences. TRAPPIST-North is a project funded by the University of Li{\`e}ge, in collaboration with Cadi Ayyad University of Marrakech (Morocco). TRAPPIST-South is a project funded by the Belgian FNRS under grant FRFC 2.5.594.09.F. E.~Jehin is a FNRS Senior Research Associate.

ASAS-SN thanks the Las Cumbres Observatory and its staff for its continuing support of the ASAS-SN project. ASAS-SN is supported by the Gordon and Betty Moore Foundation through grant GBMF5490 to the Ohio State University and NSF grant AST-1515927. Development of ASAS-SN has been supported by NSF grant AST-0908816, the Mt. Cuba Astronomical Foundation, the Center for Cosmology and AstroParticle Physics at the Ohio State University, the Chinese Academy of Sciences South America Center for Astronomy (CASSACA), the Villum Foundation, and George Skestos. 
\end{acknowledgements}

\bibliography{benoit,mybib}
\bibliographystyle{aa}

\begin{appendix}

\section{Additional figures and tables}

\setkeys{Gin}{draft=false}
\begin{figure*}
\begin{center}
\resizebox{1.0\hsize}{!}{\includegraphics{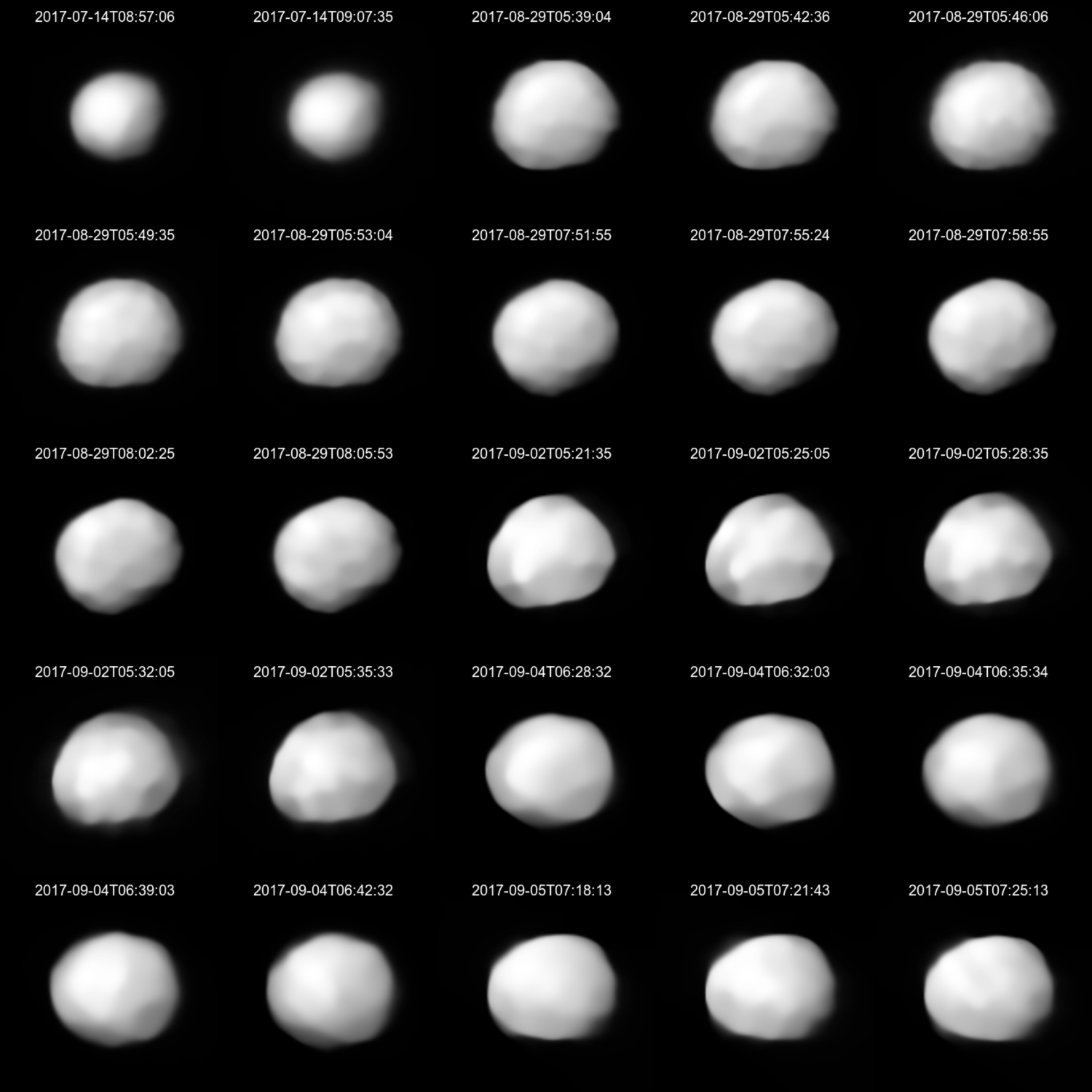}}\\
\end{center}
\caption{\label{fig:Deconv1}Full set of VLT/SPHERE/ZIMPOL images of (704) Interamnia obtained in August--September 2017. All images were deconvolved with the \mistral~algorithm. Table~\ref{tab:ao} contains full information about the data.}
\end{figure*}
\setkeys{Gin}{draft=true}

\setkeys{Gin}{draft=false}
\begin{figure*}
\begin{center}
\resizebox{0.85\hsize}{!}{\includegraphics{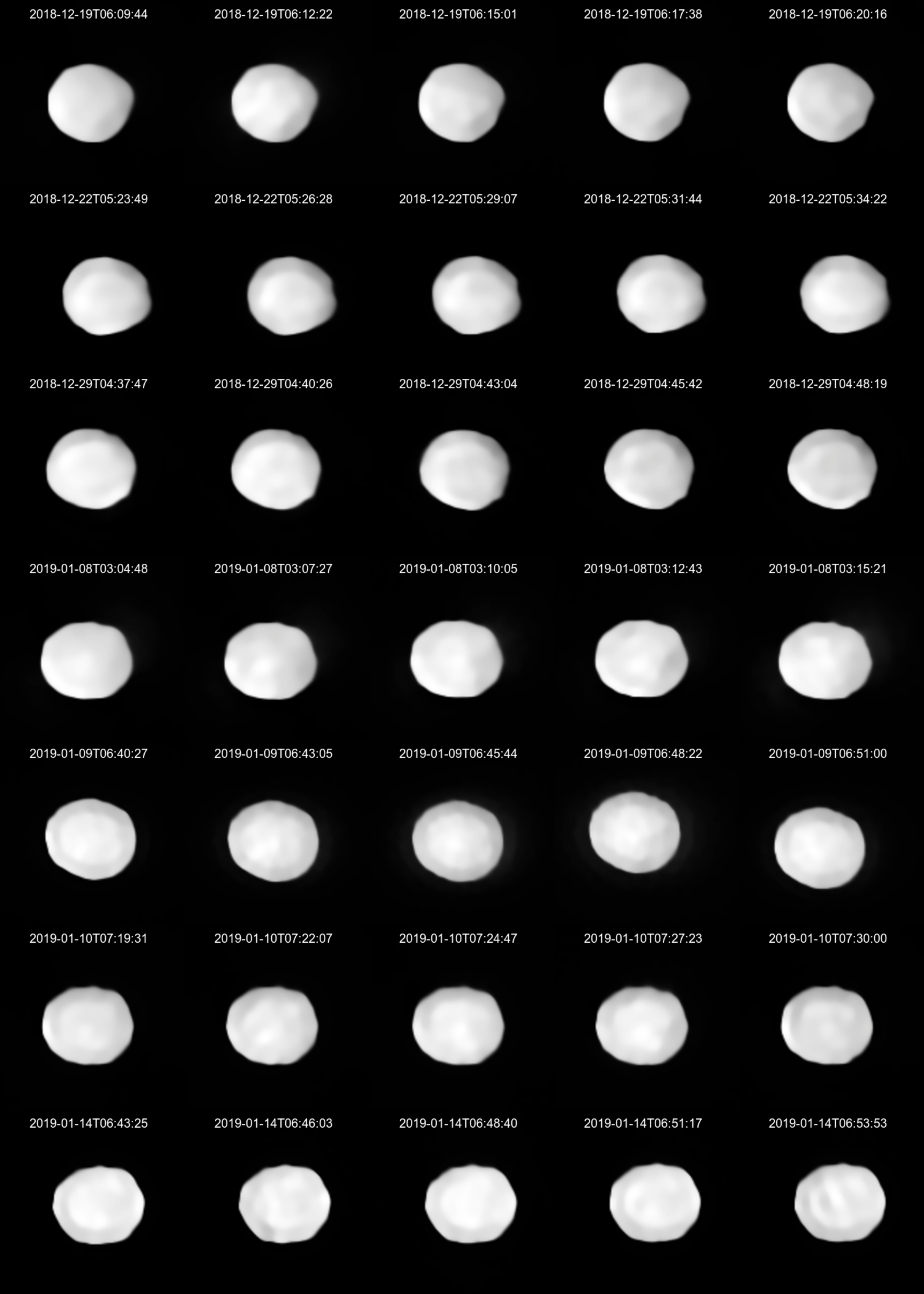}}\\
\end{center}
\caption{\label{fig:Deconv2}Full set of VLT/SPHERE/ZIMPOL images of (704) Interamnia obtained between December 2018 and January 2019. All images were deconvolved with the \mistral~algorithm. Table~\ref{tab:ao} contains full information about the data.}
\end{figure*}
\setkeys{Gin}{draft=true}

\setkeys{Gin}{draft=false}
\begin{figure}
\begin{center}
\resizebox{1.0\hsize}{!}{\includegraphics{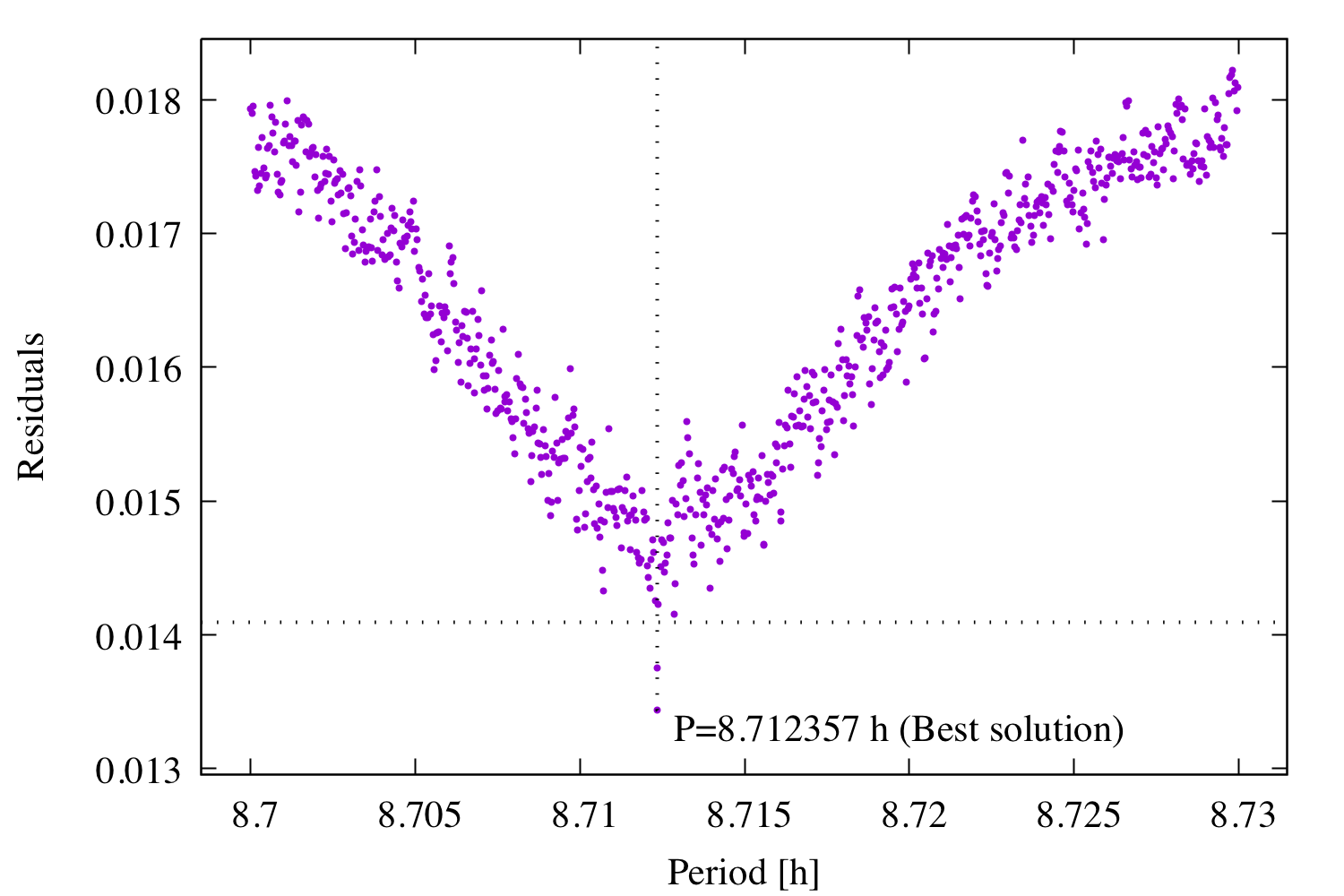}}\\
\end{center}
\caption{\label{fig:period}Periodogram for Interamnia: each point corresponds to a local minimum in the parameter space. The point with the lowest rms is the global minimum and the horizontal line indicates the rms threshold as defined in \citet{Hanus2018c}.}
\end{figure}
\setkeys{Gin}{draft=true}

\setkeys{Gin}{draft=false}
\begin{figure*}
\begin{center}
\resizebox{1.0\hsize}{!}{\includegraphics{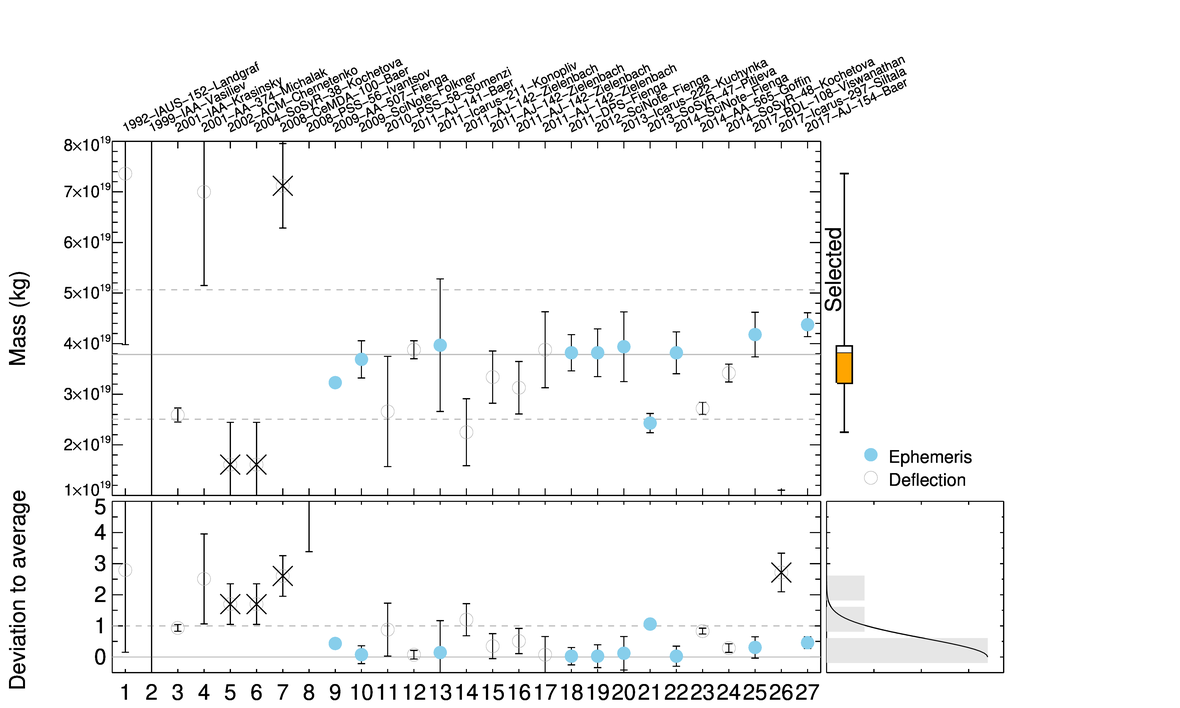}}\\
\end{center}
\caption{\label{fig:mass}Mass estimates ($\mathcal{M}$) of (704)~Interamnia collected in the literature.}
\end{figure*}
\setkeys{Gin}{draft=true}

\setkeys{Gin}{draft=false}
\begin{figure}
\begin{center}
\resizebox{0.90\hsize}{!}{\includegraphics{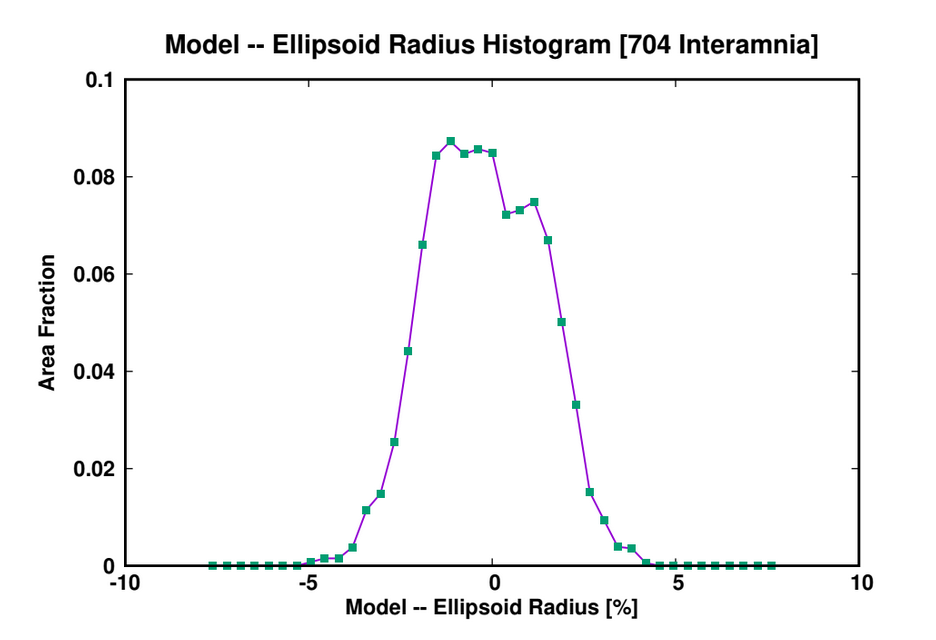}}\\
\end{center}
\caption{\label{fig:histogram}Distribution of residuals measured along the local normal direction between the \adam{} shape model and the best-fitting ellipsoid.}
\end{figure}
\setkeys{Gin}{draft=true}


\begin{table*}
\caption{\label{tab:ao}List of VLT/SPHERE disk-resolved images obtained in the I filter by the ZIMPOL camera. For each observation, the table gives the epoch, the exposure time, the airmass, the distance to the Earth $\Delta$ and the Sun $r$, the phase angle $\alpha$, and the angular diameter $D_\mathrm{a}$.}
\centering
\begin{tabular}{rr rr rrr r}
\hline 
\multicolumn{1}{c} {Date} & \multicolumn{1}{c} {UT} & \multicolumn{1}{c} {Exp} & \multicolumn{1}{c} {Airmass} & \multicolumn{1}{c} {$\Delta$} & \multicolumn{1}{c} {$r$} & \multicolumn{1}{c} {$\alpha$} & \multicolumn{1}{c} {$D_\mathrm{a}$} \\
\multicolumn{1}{c} {} & \multicolumn{1}{c} {} & \multicolumn{1}{c} {(s)} & \multicolumn{1}{c} {} & \multicolumn{1}{c} {(AU)} & \multicolumn{1}{c} {(AU)} & \multicolumn{1}{c} {(\degr)} & \multicolumn{1}{c} {(\arcsec)} \\
\hline\hline
  2017-07-14 &      8:57:06 & 200 & 1.46 & 2.24 & 2.62 & 22.4 & 0.203 \\
  2017-07-14 &      9:07:35 & 200 & 1.45 & 2.24 & 2.62 & 22.4 & 0.203 \\
  2017-08-29 &      5:39:04 & 200 & 1.71 & 1.78 & 2.60 & 15.9 & 0.256 \\
  2017-08-29 &      5:42:36 & 200 & 1.70 & 1.78 & 2.60 & 15.9 & 0.256 \\
  2017-08-29 &      5:46:06 & 200 & 1.70 & 1.78 & 2.60 & 15.9 & 0.256 \\
  2017-08-29 &      5:49:35 & 200 & 1.69 & 1.78 & 2.60 & 15.9 & 0.256 \\
  2017-08-29 &      5:53:04 & 200 & 1.69 & 1.78 & 2.60 & 15.9 & 0.256 \\
  2017-08-29 &      7:51:55 & 200 & 1.88 & 1.78 & 2.60 & 15.9 & 0.256 \\
  2017-08-29 &      7:55:24 & 200 & 1.89 & 1.78 & 2.60 & 15.9 & 0.256 \\
  2017-08-29 &      7:58:55 & 200 & 1.91 & 1.78 & 2.60 & 15.9 & 0.256 \\
  2017-08-29 &      8:02:25 & 200 & 1.93 & 1.78 & 2.60 & 15.9 & 0.256 \\
  2017-08-29 &      8:05:53 & 200 & 1.95 & 1.78 & 2.60 & 15.9 & 0.256 \\
  2017-09-02 &      5:21:35 & 200 & 1.72 & 1.76 & 2.60 & 15.0 & 0.259 \\
  2017-09-02 &      5:25:05 & 200 & 1.71 & 1.76 & 2.60 & 15.0 & 0.259 \\
  2017-09-02 &      5:28:35 & 200 & 1.71 & 1.76 & 2.60 & 15.0 & 0.259 \\
  2017-09-02 &      5:32:05 & 200 & 1.70 & 1.76 & 2.60 & 15.0 & 0.259 \\
  2017-09-02 &      5:35:33 & 200 & 1.70 & 1.76 & 2.60 & 15.0 & 0.259 \\
  2017-09-04 &      6:28:32 & 200 & 1.72 & 1.74 & 2.60 & 14.5 & 0.261 \\
  2017-09-04 &      6:32:03 & 200 & 1.73 & 1.74 & 2.60 & 14.5 & 0.261 \\
  2017-09-04 &      6:35:34 & 200 & 1.73 & 1.74 & 2.60 & 14.5 & 0.261 \\
  2017-09-04 &      6:39:03 & 200 & 1.74 & 1.74 & 2.60 & 14.5 & 0.261 \\
  2017-09-04 &      6:42:32 & 200 & 1.75 & 1.74 & 2.60 & 14.5 & 0.261 \\
  2017-09-05 &      7:18:13 & 200 & 1.89 & 1.74 & 2.60 & 14.3 & 0.261 \\
  2017-09-05 &      7:21:43 & 200 & 1.91 & 1.74 & 2.60 & 14.3 & 0.261 \\
  2017-09-05 &      7:25:13 & 200 & 1.92 & 1.74 & 2.60 & 14.3 & 0.261 \\
  2018-12-19 &      6:09:44 & 147 & 1.35 & 2.15 & 3.02 & 10.1 & 0.212 \\
  2018-12-19 &      6:12:22 & 147 & 1.35 & 2.15 & 3.02 & 10.1 & 0.212 \\
  2018-12-19 &      6:15:01 & 147 & 1.34 & 2.15 & 3.02 & 10.1 & 0.212 \\
  2018-12-19 &      6:17:38 & 147 & 1.34 & 2.15 & 3.02 & 10.1 & 0.212 \\
  2018-12-19 &      6:20:16 & 147 & 1.34 & 2.15 & 3.02 & 10.1 & 0.212 \\
  2018-12-22 &      5:23:49 & 147 & 1.41 & 2.13 & 3.03 & 9.2 & 0.214 \\
  2018-12-22 &      5:26:28 & 147 & 1.40 & 2.13 & 3.03 & 9.2 & 0.214 \\
  2018-12-22 &      5:29:07 & 147 & 1.39 & 2.13 & 3.03 & 9.2 & 0.214 \\
  2018-12-22 &      5:31:44 & 147 & 1.39 & 2.13 & 3.03 & 9.2 & 0.214 \\
  2018-12-22 &      5:34:22 & 147 & 1.38 & 2.13 & 3.03 & 9.2 & 0.214 \\
  2018-12-29 &      4:37:47 & 147 & 1.43 & 2.10 & 3.04 & 6.7 & 0.217 \\
  2018-12-29 &      4:40:26 & 147 & 1.43 & 2.10 & 3.04 & 6.7 & 0.217 \\
  2018-12-29 &      4:43:04 & 147 & 1.42 & 2.10 & 3.04 & 6.7 & 0.217 \\
  2018-12-29 &      4:45:42 & 147 & 1.41 & 2.10 & 3.04 & 6.7 & 0.217 \\
  2018-12-29 &      4:48:19 & 147 & 1.40 & 2.10 & 3.04 & 6.7 & 0.217 \\
  2019-01-08 &      3:04:48 & 147 & 1.60 & 2.08 & 3.05 & 3.3 & 0.219 \\
  2019-01-08 &      3:07:27 & 147 & 1.59 & 2.08 & 3.05 & 3.3 & 0.219 \\
  2019-01-08 &      3:10:05 & 147 & 1.58 & 2.08 & 3.05 & 3.3 & 0.219 \\
  2019-01-08 &      3:12:43 & 147 & 1.56 & 2.08 & 3.05 & 3.3 & 0.219 \\
  2019-01-08 &      3:15:21 & 147 & 1.55 & 2.08 & 3.05 & 3.3 & 0.219 \\
  2019-01-09 &      6:40:27 & 147 & 1.42 & 2.08 & 3.05 & 2.9 & 0.219 \\
  2019-01-09 &      6:43:05 & 147 & 1.43 & 2.08 & 3.05 & 2.9 & 0.219 \\
  2019-01-09 &      6:45:44 & 147 & 1.44 & 2.08 & 3.05 & 2.9 & 0.219 \\
  2019-01-09 &      6:48:22 & 147 & 1.44 & 2.08 & 3.05 & 2.9 & 0.219 \\
  2019-01-09 &      6:51:00 & 147 & 1.45 & 2.08 & 3.05 & 2.9 & 0.219 \\
  2019-01-10 &      7:19:31 & 147 & 1.59 & 2.08 & 3.06 & 2.6 & 0.219 \\
  2019-01-10 &      7:22:07 & 147 & 1.61 & 2.08 & 3.06 & 2.6 & 0.219 \\
  2019-01-10 &      7:24:47 & 147 & 1.62 & 2.08 & 3.06 & 2.6 & 0.219 \\
  2019-01-10 &      7:27:23 & 147 & 1.64 & 2.08 & 3.06 & 2.6 & 0.219 \\
  2019-01-10 &      7:30:00 & 147 & 1.65 & 2.08 & 3.06 & 2.6 & 0.219 \\
  2019-01-14 &      6:43:25 & 147 & 1.51 & 2.08 & 3.06 & 2.0 & 0.219 \\
  2019-01-14 &      6:46:03 & 147 & 1.52 & 2.08 & 3.06 & 2.0 & 0.219 \\
  2019-01-14 &      6:48:40 & 147 & 1.53 & 2.08 & 3.06 & 2.0 & 0.219 \\
  2019-01-14 &      6:51:17 & 147 & 1.55 & 2.08 & 3.06 & 2.0 & 0.219 \\
  2019-01-14 &      6:53:53 & 147 & 1.56 & 2.08 & 3.06 & 2.0 & 0.219 \\
\hline
\end{tabular}
\end{table*}

\onecolumn
\begin{longtable}{l}
\caption{\label{tab:occ}List of stellar occultations used for shape modeling, with individual observers names.}\\
\hline
\multicolumn{1}{c} {\textbf{(704) Interamnia}} \\ \hline\hline

\endfirsthead
\caption{continued.}\\

\hline
 Observer \\ \hline\hline
\endhead
\hline
\endfoot
\multicolumn{1}{c} {\textbf{1996-12-17}} \\
Bob Fried, Braeside Obs.,Flagstaff, AZ, USA \\
Gary Goodman, Camarillo, CA, USA           \\
Etscorn Obs., Socorro, NM, USA             \\
Orange County A.S. Obs., Anza, CA, USA\\
F. Wright/Fulton Jr., Prescott, AZ, USA \\
Pierre Schwaar, Phoenix, AZ, USA    \\
P. Maley/L Paller, Phoenix, AZ, USA    \\
Sam Herchak, Mesa, AZ, USA         \\
Table Mtn. Obs., Wrightwood, CA, USA   \\
Ken Ziegler, Gila Obs., Globe, AZ, USA \\
Bill Peters, AZ, USA                   \\ \hline
\multicolumn{1}{c} {\textbf{2003-3-23}} \\
Yoshida Hidetoshi, Abashiri, Hokkaido, JP \\
Kouda Masaki, Kamikita, Aomori, JP    \\
Imatani \& Takashi, Kitaura, Ibaraka, JP \\
Sugawara Hitoshi, Ichinoseki, Iwate, JP \\
Satou Toshirou, Ichinoseki, Iwate, JP \\
Yokokawa Mikio, Motoyoshi, Miyagi, JP \\
Konno Eitoshi, Hanaizumi, Iwate JP    \\
Sasaki Kazuo, Furukawa, Miyagi, JP    \\
Tonomura Yasuhiro, Tomiya, Miyagi, JP \\
Okamoto Michiko, Rifu, Miyagi, JP     \\
Sakaki Chiyoaki, Sendai, Miyagi, JP   \\
Nagai Hideo, Sendai, Miyagi, JP       \\
Itou Yoshiharu, Aoba, Sendai, JP      \\
Ikeshita Ryo et al., Kawauchi, Sendai, JP \\
Koishikawa Masahiro, Sendai, Miyagi, JP \\
Watanabe, Akira, Sendai, Miyagi, JP   \\
Miyamoto Atsushi et al, Adachi, JP    \\
Sugai Hideo, Zao-hango, Yamagata, JP  \\
Fujita Mitsuhiro, Shiroishi, Miyagi, JP \\
Nihei Hajime, Nanyo, Yamagata, JP     \\
Tanaka Takashi, Zushi, Kanagawa, JP    \\
Sato Tsutomu, Marumori, Miyagi, JP    \\
Ootsuki Isao, Marumori, Miyagi, JP    \\
David Tholen, Turtle Bay, Oahu, Hawaii, USA \\
Sato Hikaru, Fukushima, JP            \\
Sato Makoto, Haranomachi, Fukushima, JP \\
Rebecca Sydney, Honolua Bay, Maui, HI \\
Bedient et al., Foster Village, Hawaii, USA \\
Hamanowa et al., Koriyama City, JP     \\
Usuki Ken-ichi, Niitsuru, Fukushima, JP \\
Lewis Roberts, Haleakala, Hawaii, USA  \\
Watanabe et al., Takine, Fukushima, JP \\
Sato Hirohisa, Sukagawa, Fukushima, JP \\
David Dunham, Makena, Maui, HI, USA    \\
B. Brevoort, Hawaii, USA                \\
Tsuchikawa Akira, Yanagida, JP        \\
S. Bus, Mauna Kea, Hawaii, USA         \\
R. Savalle, Mauna Kea, Hawaii, USA     \\
Tomioka Hiroyuki, Hitachi, Ibaraki, JP \\
P. Maley, Hawaii, USA                  \\
Yaeza Akira,Moriyama-cho, Hitachi, JP  \\
E. Cleintuar, Hawaii, USA              \\
S. O'Meara, Mauna Loa, USA             \\
W. Fukunaga, Hawaii, USA               \\
V. Fukunaga, Hawaii, USA               \\
J. Swatek, Hawaii, USA                 \\
Uehara Sadaharu, Ibaraki, JP          \\
Kuboniwa Atasushi, Ushika,Ibaraka, JP \\
Kita Nobusuke, Kashiwa, Chiba, JP     \\
Takashima et al, Kashiwa, Chiba, JP   \\
Momose Masahiko, Shiojiri, Nagano, JP \\
Kaneko Sakae, Sakura, Chiba, JP       \\
Nakanishi Akio, Itabashi, Tokyo, JP   \\
Ishida Masayuki, Kanazu, Fukui, JP    \\
Kitazaki Katsuhiko, Tokyo, JP         \\
Ida Miyoshi, Muraoka, Fukui, JP       \\
Suzuki Satoshi, Yokohama,Kanagawa, JP \\
Hirose Yoji, Chigasaki, Kanagawa, JP  \\
Sugiyama Yukihiro, Hiratsuka, JP      \\
Yoneyama Seiichi, Ogaki, Gifu, JP     \\
Oribe Takaaki et al., Saji, Tottori, JP \\ \hline
\multicolumn{1}{c} {\textbf{2007-9-9}} \\
R. Cadmus, Grinnell, IA, USA          \\
J. Centala, Marion, IA, USA            \\
W. Osborn/C. Tycner, Mt. Pleasant, MI, USA\\
P. Maley, Bernalillo, NM, USA          \\
K. McKeown, Los Lunas, NM, USA          \\
D. Dunham, Dubuque IA, USA             \\
\multicolumn{1}{c} {\textbf{2012-11-12}} \\ \hline
N. Smith, Trenton, GA, USA               \\
T. Blank/M. Pacht, Taftsville, VT, USA   \\
S. Conard, Gamber, MD, USA            \\
A. Scheck, Scaggsville, MD, USA        \\
C. Ellington, Owings, MD, USA           \\
\hline
 \end{longtable}

\onecolumn
\begin{longtable}{rlr rrr l l}
\caption{\label{tab:lcs}List of optical disk-integrated lightcurves used for \adam{} shape modeling. For each lightcurve, the table gives the epoch, the number of individual measurements $N_p$, asteroid's distances to the Earth $\Delta$ and the Sun $r$, phase angle $\varphi$, photometric filter and observation information.}\\
\hline 
\multicolumn{1}{c} {N} & \multicolumn{1}{c} {Epoch} & \multicolumn{1}{c} {$N_p$} & \multicolumn{1}{c} {$\Delta$} & \multicolumn{1}{c} {$r$} & \multicolumn{1}{c} {$\varphi$} & \multicolumn{1}{c} {Filter} & Reference \\
 &  &  & (AU) & (AU) & (\degr) &  &  \\
\hline\hline
\endfirsthead
\caption{continued.}\\
\hline
\multicolumn{1}{c} {N} & \multicolumn{1}{c} {Epoch} & \multicolumn{1}{c} {$N_p$} & \multicolumn{1}{c} {$\Delta$} & \multicolumn{1}{c} {$r$} & \multicolumn{1}{c} {$\varphi$} & \multicolumn{1}{c} {Filter} & Reference \\
 &  &  & (AU) & (AU) & (\degr) &  &  \\
\hline\hline
\endhead
\hline
\endfoot
\hline
     1    &  1964-11-22.7  &  81    &  1.82  &  2.75  &  8.1   &  V &  \citet{Yang1965} \\
     2    &  1964-11-23.7  &  57    &  1.81  &  2.75  &  7.7   &  V &  \citet{Yang1965} \\
     3    &  1969-08-16.5  &  8     &  2.18  &  2.58  &  22.6  &  V &  \citet{Tempesti1975} \\
     4    &  1969-08-20.6  &  7     &  2.14  &  2.58  &  22.3  &  V &  \citet{Tempesti1975} \\
     5    &  1969-08-21.5  &  12    &  2.13  &  2.58  &  22.2  &  V &  \citet{Tempesti1975} \\
     6    &  1969-09-08.5  &  12    &  1.95  &  2.59  &  19.9  &  V &  \citet{Tempesti1975} \\
     7    &  1969-10-05.6  &  13    &  1.74  &  2.60  &  13.9  &  V &  \citet{Tempesti1975} \\
     8    &  1969-10-15.4  &  10    &  1.70  &  2.60  &  11.4  &  V &  \citet{Tempesti1975} \\
     9    &  1969-11-01.4  &  11    &  1.68  &  2.61  &  9.1   &  V &  \citet{Tempesti1975} \\
    10    &  1969-11-03.4  &  15    &  1.68  &  2.61  &  9.2   &  V &  \citet{Tempesti1975} \\
    11    &  1969-11-10.5  &  23    &  1.69  &  2.62  &  9.8   &  V &  \citet{Tempesti1975} \\
    12    &  1974-08-27.1  &  29    &  1.71  &  2.62  &  12.2  &  V &  \citet{Lustig1976}    \\
    13    &  1974-08-28.0  &  91    &  1.71  &  2.62  &  12.1  &  V &  \citet{Lustig1976}  \\
    14    &  1974-08-29.0  &  92    &  1.71  &  2.62  &  11.9  &  V &  \citet{Lustig1976}  \\
    15    &  1990-08-01.9  &  76    &  1.86  &  2.68  &  15.5  &  V &  \citet{Shevchenko1992} \\
    16    &  1993-03-21.6  &  77    &  2.55  &  3.47  &  7.6   &  V &  \citet{Michalowski1995b} \\
    17    &  1993-03-23.6  &  73    &  2.56  &  3.47  &  7.9   &  V &  \citet{Michalowski1995b} \\
    18    &  1996-12-13.2  &  512   &  1.83  &  2.74  &  9.6   &  V &  \citet{Sato2014}     \\
    19    &  1996-12-13.3  &  354   &  1.83  &  2.74  &  9.6   &  R &  \citet{Sato2014}     \\
    20    &  2003-03-31.9  &  29    &  2.85  &  3.20  &  17.8  &  C &  Stephane Charbonnel     \\
    21    &  2003-03-31.9  &  55    &  2.85  &  3.20  &  17.8  &  C &  Nathanal Berger  \\
    22    &  2006-06-04.1  &  30    &  2.46  &  2.79  &  21.0  &  C &  Arnaud Leroy     \\
    23    &  2006-07-21    &  59    &  1.89  &  2.74  &  14.3  &  C &  \citet{Grice2017}     \\
    24    &  2006-07-25    &  53    &  1.86  &  2.73  &  13.3  &  C &  \citet{Grice2017}  \\
    25    &  2006-07-25    &  54    &  1.86  &  2.73  &  13.3  &  C &  \citet{Grice2017}  \\
    26    &  2006-07-26    &  38    &  1.85  &  2.73  &  13.1  &  C &  \citet{Grice2017}\\
    27    &  2006-07-26    &  40    &  1.85  &  2.73  &  13.1  &  C &  \citet{Grice2017}\\
    28    &  2006-07-27    &  50    &  1.85  &  2.73  &  12.8  &  C &  \citet{Grice2017}   \\
    29    &  2006-07-27    &  55    &  1.85  &  2.73  &  12.8  &  C &  \citet{Grice2017}   \\
    30    &  2006-07-28    &  56    &  1.84  &  2.73  &  12.6  &  C &  \citet{Grice2017}   \\
    31    &  2006-07-28    &  56    &  1.84  &  2.73  &  12.6  &  C &  \citet{Grice2017}   \\
    32    &  2006-07-29    &  54    &  1.83  &  2.73  &  12.3  &  C &  \citet{Grice2017}   \\
    33    &  2006-07-29    &  58    &  1.83  &  2.73  &  12.3  &  C &  \citet{Grice2017}   \\
    34    &  2006-07-30    &  59    &  1.82  &  2.72  &  12.1  &  C &  \citet{Grice2017}   \\
    35    &  2006-07-30    &  59    &  1.82  &  2.72  &  12.1  &  C &  \citet{Grice2017}   \\
    36    &  2006-07-31    &  46    &  1.82  &  2.72  &  11.8  &  C &  \citet{Grice2017}   \\
    37    &  2006-07-31    &  59    &  1.82  &  2.72  &  11.8  &  C &  \citet{Grice2017}   \\
    38    &  2006-08-02    &  64    &  1.81  &  2.72  &  11.3  &  C &  \citet{Grice2017}   \\
    39    &  2006-08-04    &  89    &  1.79  &  2.72  &  10.8  &  C &  \citet{Grice2017}   \\
    40    &  2006-08-05    &  51    &  1.79  &  2.72  &  10.6  &  C &  \citet{Grice2017}   \\
    41    &  2006-08-05    &  87    &  1.79  &  2.72  &  10.6  &  C &  \citet{Grice2017}   \\
    42    &  2006-08-06    &  61    &  1.78  &  2.72  &  10.3  &  C &  \citet{Grice2017}   \\
    43    &  2006-08-06    &  89    &  1.78  &  2.72  &  10.3  &  C &  \citet{Grice2017}   \\
    44    &  2006-08-07    &  62    &  1.78  &  2.72  &  10.1  &  C &  \citet{Grice2017}   \\
    45    &  2006-08-07    &  88    &  1.78  &  2.72  &  10.1  &  C &  \citet{Grice2017}   \\
    46    &  2006-08-12    &  105   &  1.76  &  2.71  &  9.0   &  C &  \citet{Grice2017}   \\
    47    &  2006-08-12    &  118   &  1.76  &  2.71  &  9.0   &  C &  \citet{Grice2017}   \\
    48    &  2006-08-13    &  67    &  1.75  &  2.71  &  8.9   &  C &  \citet{Grice2017}   \\
    49    &  2006-08-13    &  91    &  1.75  &  2.71  &  8.9   &  C &  \citet{Grice2017}   \\
    50    &  2006-08-19    &  49    &  1.74  &  2.70  &  8.1   &  C &  \citet{Grice2017}   \\
    51    &  2006-08-20    &  41    &  1.74  &  2.70  &  8.0   &  C &  \citet{Grice2017}   \\
    52    &  2006-08-22    &  54    &  1.73  &  2.70  &  7.9   &  C &  \citet{Grice2017}   \\
    53    &  2006-08-25    &  42    &  1.73  &  2.70  &  8.0   &  C &  \citet{Grice2017}   \\
    54    &  2006-08-30    &  52    &  1.73  &  2.69  &  8.4   &  C &  \citet{Grice2017}   \\
    55    &  2006-08-31    &  67    &  1.73  &  2.69  &  8.5   &  C &  \citet{Grice2017}   \\
    56    &  2006-08-32.0  &  236   &  1.73  &  2.69  &  8.5   &  C &  Dominique Suys, Hugo Riemis, Jan Vantomme\\
    57    &  2006-09-02    &  52    &  1.74  &  2.69  &  8.9   &  C &  \citet{Grice2017}   \\
    58    &  2006-09-12    &  57    &  1.77  &  2.68  &  11.1  &  C &  \citet{Grice2017}   \\
    59    &  2006-09-14    &  50    &  1.77  &  2.68  &  11.6  &  C &  \citet{Grice2017}   \\
    60    &  2006-09-16    &  46    &  1.78  &  2.68  &  12.1  &  C &  \citet{Grice2017}   \\
    61    &  2006-09-19    &  38    &  1.80  &  2.67  &  12.9  &  C &  \citet{Grice2017}   \\
    62    &  2006-09-22    &  46    &  1.82  &  2.67  &  13.7  &  C &  \citet{Grice2017}   \\
    63    &  2006-09-24    &  47    &  1.83  &  2.67  &  14.3  &  C &  \citet{Grice2017}   \\
    64    &  2006-09-26    &  47    &  1.84  &  2.67  &  14.8  &  C &  \citet{Grice2017}   \\
    65    &  2006-09-27    &  38    &  1.85  &  2.67  &  15.0  &  C &  \citet{Grice2017}   \\
    66    &  2006-10-08    &  43    &  1.94  &  2.66  &  17.6  &  C &  \citet{Grice2017}   \\
    67    &  2007-10-18    &  39    &  2.29  &  2.78  &  19.8  &  C &  \citet{Grice2017}   \\
    68    &  2007-11-22    &  80    &  1.96  &  2.82  &  11.7  &  C &  \citet{Grice2017}   \\
    69    &  2007-12-27.9  &  34    &  1.90  &  2.87  &  3.1   &  C &  Jean-Francois Coliac    \\
    70    &  2008-12-28    &  58    &  2.94  &  3.37  &  16.2  &  C &  \citet{Grice2017}   \\
    71    &  2009-01-03    &  59    &  2.87  &  3.37  &  15.5  &  C &  \citet{Grice2017}   \\
    72    &  2009-01-04    &  58    &  2.85  &  3.38  &  15.4  &  C &  \citet{Grice2017}   \\
    73    &  2009-01-17    &  61    &  2.71  &  3.39  &  13.5  &  C &  \citet{Grice2017}   \\
    74    &  2009-01-18    &  55    &  2.70  &  3.39  &  13.4  &  C &  \citet{Grice2017}   \\
    75    &  2009-01-20    &  53    &  2.68  &  3.39  &  13.0  &  C &  \citet{Grice2017}   \\
    76    &  2009-01-20    &  73    &  2.68  &  3.39  &  13.0  &  C &  \citet{Grice2017}   \\
    77    &  2009-01-21    &  58    &  2.67  &  3.39  &  12.8  &  C &  \citet{Grice2017}   \\
    78    &  2009-01-23    &  51    &  2.65  &  3.39  &  12.4  &  C &  \citet{Grice2017}   \\
    79    &  2009-01-24    &  55    &  2.65  &  3.39  &  12.2  &  C &  \citet{Grice2017}   \\
    80    &  2009-01-24    &  77    &  2.64  &  3.39  &  12.2  &  C &  \citet{Grice2017}   \\
    81    &  2009-01-25    &  56    &  2.64  &  3.40  &  12.0  &  C &  \citet{Grice2017}   \\
    82    &  2009-01-26    &  61    &  2.63  &  3.40  &  11.8  &  C &  \citet{Grice2017}   \\
    83    &  2009-01-27    &  71    &  2.62  &  3.40  &  11.6  &  C &  \citet{Grice2017}   \\
    84    &  2009-01-28    &  73    &  2.61  &  3.40  &  11.4  &  C &  \citet{Grice2017}   \\
    85    &  2009-01-31    &  60    &  2.59  &  3.40  &  10.8  &  C &  \citet{Grice2017}   \\
    86    &  2009-02-01    &  37    &  2.58  &  3.40  &  10.6  &  C &  \citet{Grice2017}   \\
    87    &  2009-02-02    &  77    &  2.58  &  3.40  &  10.4  &  C &  \citet{Grice2017}   \\
    88    &  2009-02-16    &  46    &  2.50  &  3.42  &  7.5   &  C &  \citet{Grice2017}   \\
    89    &  2009-02-18    &  85    &  2.50  &  3.42  &  7.1   &  C &  \citet{Grice2017}   \\
    90    &  2009-02-19    &  46    &  2.49  &  3.42  &  7.0   &  C &  \citet{Grice2017}   \\
    91    &  2009-02-21    &  81    &  2.49  &  3.42  &  6.7   &  C &  \citet{Grice2017}   \\
    92    &  2009-02-21    &  84    &  2.49  &  3.42  &  6.7   &  C &  \citet{Grice2017}   \\
    93    &  2009-02-22    &  47    &  2.49  &  3.42  &  6.6   &  C &  \citet{Grice2017}   \\
    94    &  2009-02-22    &  63    &  2.49  &  3.42  &  6.6   &  C &  \citet{Grice2017}   \\
    95    &  2009-02-26    &  40    &  2.48  &  3.42  &  6.2   &  C &  \citet{Grice2017}   \\
    96    &  2009-02-27    &  48    &  2.48  &  3.42  &  6.1   &  C &  \citet{Grice2017}   \\
    97    &  2009-03-01    &  50    &  2.48  &  3.43  &  6.1   &  C &  \citet{Grice2017}   \\
    98    &  2009-03-02    &  83    &  2.48  &  3.43  &  6.1   &  C &  \citet{Grice2017}   \\
    99    &  2009-03-04    &  55    &  2.49  &  3.43  &  6.1   &  C &  \citet{Grice2017}   \\
   100    &  2009-03-13    &  93    &  2.51  &  3.44  &  7.0   &  C &  \citet{Grice2017}   \\
   101    &  2009-03-14    &  65    &  2.51  &  3.44  &  7.2   &  C &  \citet{Grice2017}   \\
   102    &  2009-03-16    &  53    &  2.52  &  3.44  &  7.6   &  C &  \citet{Grice2017}   \\
   103    &  2009-03-17    &  61    &  2.53  &  3.44  &  7.7   &  C &  \citet{Grice2017}   \\
   104    &  2009-03-19    &  64    &  2.54  &  3.44  &  8.1   &  C &  \citet{Grice2017}   \\
   105    &  2009-03-20    &  47    &  2.54  &  3.44  &  8.3   &  C &  \citet{Grice2017}   \\
   106    &  2009-03-20    &  79    &  2.54  &  3.44  &  8.3   &  C &  \citet{Grice2017}   \\
   107    &  2009-03-21    &  42    &  2.55  &  3.44  &  8.5   &  C &  \citet{Grice2017}   \\
   108    &  2009-03-21    &  71    &  2.55  &  3.44  &  8.5   &  C &  \citet{Grice2017}   \\
   109    &  2009-03-22    &  49    &  2.55  &  3.44  &  8.8   &  C &  \citet{Grice2017}   \\
   110    &  2009-03-22    &  76    &  2.55  &  3.44  &  8.7   &  C &  \citet{Grice2017}   \\
   111    &  2009-03-23    &  65    &  2.56  &  3.44  &  8.9   &  C &  \citet{Grice2017}   \\
   112    &  2009-03-28    &  85    &  2.59  &  3.45  &  10.0  &  C &  \citet{Grice2017}   \\
   113    &  2009-03-29    &  83    &  2.60  &  3.45  &  10.2  &  C &  \citet{Grice2017}   \\
   114    &  2009-03-30    &  36    &  2.61  &  3.45  &  10.4  &  C &  \citet{Grice2017}   \\
   115    &  2009-03-30    &  74    &  2.61  &  3.45  &  10.4  &  C &  \citet{Grice2017}   \\
   116    &  2009-03-30    &  77    &  2.61  &  3.45  &  10.4  &  C &  \citet{Grice2017}   \\
   117    &  2009-03-31    &  52    &  2.62  &  3.45  &  10.6  &  C &  \citet{Grice2017}   \\
   118    &  2009-04-01    &  75    &  2.63  &  3.45  &  10.8  &  C &  \citet{Grice2017}   \\
   119    &  2009-04-01    &  77    &  2.63  &  3.45  &  10.8  &  C &  \citet{Grice2017}   \\
   120    &  2009-04-02    &  75    &  2.64  &  3.45  &  11.0  &  C &  \citet{Grice2017}   \\
   121    &  2009-04-12    &  71    &  2.73  &  3.46  &  12.9  &  C &  \citet{Grice2017}   \\
   122    &  2009-05-12    &  35    &  3.11  &  3.48  &  16.5  &  C &  \citet{Grice2017}   \\
   123    &  2009-05-12    &  35    &  3.11  &  3.48  &  16.5  &  C &  \citet{Grice2017}   \\
   124    &  2011-06-02    &  51    &  2.23  &  2.98  &  15.2  &  C &  \citet{Grice2017}   \\
   125    &  2011-06-12    &  50    &  2.12  &  2.97  &  12.8  &  C &  \citet{Grice2017}   \\
   126    &  2011-06-12    &  62    &  2.12  &  2.97  &  12.8  &  C &  \citet{Grice2017}   \\
   127    &  2011-06-13    &  57    &  2.11  &  2.97  &  12.5  &  C &  \citet{Grice2017}   \\
   128    &  2011-06-13    &  74    &  2.11  &  2.97  &  12.5  &  C &  \citet{Grice2017}   \\
   129    &  2011-06-20    &  41    &  2.04  &  2.96  &  10.4  &  C &  \citet{Grice2017}   \\
   130    &  2011-06-21    &  40    &  2.03  &  2.95  &  10.1  &  C &  \citet{Grice2017}   \\
   131    &  2011-07-08    &  39    &  1.93  &  2.93  &  4.5   &  C &  \citet{Grice2017}   \\
   132    &  2011-07-09    &  47    &  1.93  &  2.93  &  4.1   &  C &  \citet{Grice2017}   \\
   133    &  2011-07-09    &  69    &  1.93  &  2.93  &  4.2   &  C &  \citet{Grice2017}   \\
   134    &  2011-07-10    &  41    &  1.92  &  2.93  &  3.8   &  C &  \citet{Grice2017}   \\
   135    &  2011-07-10    &  79    &  1.92  &  2.93  &  3.8   &  C &  \citet{Grice2017}   \\
   136    &  2011-07-18    &  41    &  1.90  &  2.91  &  2.5   &  C &  \citet{Grice2017}   \\
   137    &  2011-07-31    &  39    &  1.91  &  2.89  &  6.0   &  C &  \citet{Grice2017}   \\
   138    &  2011-08-01    &  39    &  1.91  &  2.89  &  6.4   &  C &  \citet{Grice2017}   \\
   139    &  2011-08-06    &  45    &  1.93  &  2.89  &  8.1   &  C &  \citet{Grice2017}   \\
   140    &  2012-09-17.0  &  151   &  2.05  &  2.62  &  20.6  &  C &  K. Sobkowiak, Borowiec, Poland    \\
   141    &  2012-09-18.9  &  17    &  2.03  &  2.62  &  20.3  &  C &  J. Nadolny, Borowiec, Poland    \\
   142    &  2012-09-24.0  &  10    &  1.98  &  2.62  &  19.4  &  C &  J. Nadolny, Borowiec, Poland    \\
   143    &  2012-10-01.0  &  83    &  1.91  &  2.63  &  18.1  &  C &  M. Bronikowska, Borowiec, Poland    \\
   144    &  2012-11-29.9  &  45    &  1.75  &  2.67  &  9.3   &  C &  Francisco Soldan \\
   145    &  2012-12-01.9  &  48    &  1.76  &  2.67  &  9.8   &  C &  Francisco Soldan \\
   146    &  2012-12-02.9  &  59    &  1.77  &  2.68  &  10.1  &  C &  Francisco Soldan \\
   147    &  2012-12-08.9  &  29    &  1.81  &  2.68  &  11.7  &  C &  Francisco Soldan \\
   148    &  2012-12-26.0  &  18    &  1.96  &  2.70  &  16.1  &  C &  Francisco Soldan \\
   149    &  2012-12-26.0  &  24    &  1.96  &  2.70  &  16.1  &  C &  Francisco Soldan \\
   150    &  2012-12-26.8  &  174   &  1.97  &  2.70  &  16.3  &  C &  Francisco Soldan \\
   151    &  2012-12-27.8  &  145   &  1.98  &  2.70  &  16.6  &  C &  Francisco Soldan \\
   152    &  2012-12-28.8  &  23    &  1.99  &  2.70  &  16.8  &  C &  Francisco Soldan \\
   153    &  2012-12-28.8  &  55    &  1.99  &  2.70  &  16.8  &  C &  Francisco Soldan \\
   154    &  2016-05-23.1  &  52    &  2.27  &  3.18  &  9.3   &  C &  Raul Melia, Cordoba, Argentina     \\
   155    &  2016-06-15.1  &  52    &  2.14  &  3.15  &  2.3   &  C &  Carlos Colazo, Cordoba, Argentina     \\
   156    &  2016-06-16.2  &  42    &  2.14  &  3.15  &  2.1   &  C &  Carlos Colazo, Cordoba, Argentina    \\
   157    &  2016-07-25.9  &  51    &  2.26  &  3.09  &  12.9  &  C &  A. Marciniak, Obs. del Teide, Spain   \\
   158    &  2017-07-16.4  &  149   &  2.22  &  2.62  &  22.3  &  V &  \citet{Warner2018}     \\
   159    &  2017-07-17.4  &  172   &  2.21  &  2.62  &  22.2  &  V &  \citet{Warner2018}    \\
   160    &  2017-07-18.4  &  156   &  2.20  &  2.62  &  22.2  &  V &  \citet{Warner2018}    \\
   161    &  2017-07-21.4  &  201   &  2.16  &  2.62  &  22.0  &  V &  \citet{Warner2018}  \\
   162    &  2017-07-22.4  &  196   &  2.15  &  2.62  &  21.9  &  V &  \citet{Warner2018}  \\
   163    &  2017-07-23.0  &  29    &  2.14  &  2.62  &  21.8  &  r' & R. Szakats, Piszkes Obs., Hungary     \\
   164    &  2017-08-01.0  &  47    &  2.04  &  2.61  &  20.9  &  C &  Adrian Jones, Gaia-GOSA \\
   165    &  2017-08-12.0  &  23    &  1.93  &  2.61  &  19.3  &  R &  R. Szakats, Piszkes Obs., Hungary\\
   166    &  2017-08-14.0  &  35    &  1.91  &  2.61  &  19.0  &  R &  R. Szakats, Piszkes Obs., Hungary    \\
   167    &  2017-08-15.1  &  17    &  1.90  &  2.60  &  18.8  &  R &  R. Szakats, Piszkes Obs., Hungary     \\
   168    &  2017-08-16.0  &  66    &  1.89  &  2.60  &  18.6  &  R &  R. Szakats, Piszkes Obs., Hungary     \\
   169    &  2017-10-11.9  &  38    &  1.67  &  2.59  &  11.1  &  r' & R. Szakats, Piszkes Obs., Hungary  \\
   170    &  2017-10-16.8  &  139   &  1.69  &  2.59  &  11.8  &  R &  R. Duffard, La Sagra, Spain     \\
   171    &  2017-11-08.9  &  154   &  1.83  &  2.59  &  16.8  &  R &  R. Duffard, La Sagra, Spain     \\
   172    &  2017-11-15.9  &  276   &  1.89  &  2.59  &  18.2  &  R &  R. Duffard, La Sagra, Spain     \\
   173    &  2017-11-16.9  &  188   &  1.90  &  2.59  &  18.4  &  R &  R. Duffard, La Sagra, Spain     \\
   174    &  2017-12-4.0   &  76    &  2.07  &  2.59  &  20.9  &  C &  D. Molina, Gaia-GOSA    \\
   175    &  2017-7-14.1   &  58    &  2.25  &  2.62  &  22.4  &  C &  D. Molina, Gaia-GOSA    \\
   176    &  2017-7-16.2   &  63    &  2.22  &  2.62  &  22.3  &  C &  D. Molina, Gaia-GOSA    \\
   177    &  2017-7-23.2   &  48    &  2.14  &  2.62  &  21.8  &  C &  D. Molina, Gaia-GOSA    \\
   178    &  2017-7-26.2   &  35    &  2.11  &  2.62  &  21.6  &  C &  D. Molina, Gaia-GOSA    \\
   179    &  2017-7-27.2   &  58    &  2.10  &  2.62  &  21.5  &  C &  D. Molina, Gaia-GOSA    \\
   180    &  2017-8-6.1    &  187   &  1.99  &  2.61  &  20.3  &  C &  A. Jones, Gaia-GOSA    \\
   181    &  2018-12-09.1  &  83    &  2.22  &  3.01  &  13.2  &  V &  C. Garcia, Gaia-GOSA \\
   182    &  2018-12-10.1  &  100   &  2.21  &  3.01  &  13.0  &  V &  C. Garcia, Gaia-GOSA \\
   183    &  2018-12-14.0  &  54    &  2.18  &  3.01  &  11.8  &  V &  C. Garcia, Gaia-GOSA \\
   184    &  2018-12-3.3   &  1395  &  2.28  &  3.00  &  14.8  & Rc &  E. Jehin, M. Ferrais, TRAPPIST-N and -S \\
   185    &  2019-1-16.2   &  540   &  2.09  &  3.06  &  2.1   & Rc &  E. Jehin, M. Ferrais, TRAPPIST-S \\
   186    &  2019-1-19.2   &  492   &  2.09  &  3.07  &  2.7   & Rc &  E. Jehin, M. Ferrais, TRAPPIST-S \\
   187    &  2019-1-26.3   &  811   &  2.12  &  3.08  &  4.9   & Rc &  E. Jehin, M. Ferrais, TRAPPIST-S \\
   188    &  2019-2-1.2    &  269   &  2.16  &  3.09  &  7.0   & Rc &  E. Jehin, M. Ferrais, TRAPPIST-S \\
   189    &  2019-1-16.1   &  269   &  2.09  &  3.06  &  2.1   & r' &  M. Person, T. Brothers\\
   190    &  2012-12 -- 2018-1 & 198&        &        &        & V  &  ASAS-SN \\
   191    &  2015-1 -- 2016-4  & 16 &        &        &        & V  &  Gaia DR2 \\
   \hline
\end{longtable}
\tablefoot{
     Gaia-GOSA (Gaia-Ground-based Observational Service for Asteroids, \url{www.gaiagosa.eu}).
    }

\begin{table*}[h]
\begin{center}
  \caption[Mass estimates of (704) Interamnia]{
    Mass estimates ($\mathcal{M}$) of (704) Interamnia collected in the literature. For each, the 1\,$\sigma$ uncertainty, method, selection flag, and bibliographic reference are reported. The methods are \textsc{defl}: Deflection, \textsc{ephem}: Ephemeris. \label{tab:mass}}
  \begin{tabular}{rrlcl}
    \hline\hline
     \multicolumn{1}{c}{\#} & \multicolumn{1}{c}{Mass ($\mathcal{M}$)} &

     \multicolumn{1}{c}{Method} & \multicolumn{1}{c}{Sel.} & \multicolumn{1}{c}{Reference}  \\
    & \multicolumn{1}{c}{(kg)} \\
    \hline
  1 & $(7.36 \pm 3.38) \times 10^{19}$                   & \textsc{defl}  & \ding{51} & \citet{Landgraf1992}           \\
  2 & $12.3_{-12.3}^{+13.1} \times 10^{19}$              & \textsc{defl}  & \ding{55} & \citet{1999-IAA-Vasiliev}                \\
  3 & $(2.59 \pm 0.14) \times 10^{19}$                   & \textsc{defl}  & \ding{51} & \citet{2001-IAA-Krasinsky}               \\
  4 & $(7.00 \pm 1.85) \times 10^{19}$                   & \textsc{defl}  & \ding{51} & \citet{Michalak2001}             \\
  5 & $(1.61 \pm 0.84) \times 10^{19}$                  & \textsc{defl}  & \ding{55} & \citet{2002-ACM-Chernetenko}             \\
  6 & $(1.61 \pm 0.84) \times 10^{19}$                  & \textsc{defl}  & \ding{55} & \citet{2004-SoSyR-38-Kochetova}          \\
  7 & $(7.12 \pm 0.84) \times 10^{19}$                   & \textsc{defl}  & \ding{55} & \citet{Baer2008}              \\
  8 & $(11.30 \pm 3.18) \times 10^{19}$                  & \textsc{defl}  & \ding{55} & \citet{2008-PSS-56-Ivantsov}             \\
  9 & $(3.23 \pm 0.02) \times 10^{19}$                   & \textsc{ephem} & \ding{51} & \citet{Fienga2009}               \\
 10 & $(3.69 \pm 0.37) \times 10^{19}$                   & \textsc{ephem} & \ding{51} & \citet{2009-SciNote-Folkner}             \\
 11 & $(2.66 \pm 1.09) \times 10^{19}$                   & \textsc{defl}  & \ding{51} & \citet{Somenzi2010}              \\
 12 & $(3.88 \pm 0.18) \times 10^{19}$                   & \textsc{defl}  & \ding{51} & \citet{2011-AJ-141-Baer}                 \\
 13 & $(3.97 \pm 1.31) \times 10^{19}$                   & \textsc{ephem} & \ding{51} & \citet{2011-Icarus-211-Konopliv}         \\
 14 & $(2.25 \pm 0.66) \times 10^{19}$                   & \textsc{defl}  & \ding{51} & \citet{2011-AJ-142-Zielenbach}           \\
 15 & $(3.34 \pm 0.52) \times 10^{19}$                   & \textsc{defl}  & \ding{51} & \citet{2011-AJ-142-Zielenbach}           \\
 16 & $(3.13 \pm 0.52) \times 10^{19}$                   & \textsc{defl}  & \ding{51} & \citet{2011-AJ-142-Zielenbach}           \\
 17 & $(3.88 \pm 0.75) \times 10^{19}$                   & \textsc{defl}  & \ding{51} & \citet{2011-AJ-142-Zielenbach}           \\
 18 & $(3.82 \pm 0.36) \times 10^{19}$                   & \textsc{ephem} & \ding{51} & \citet{2011-DPS-Fienga}                  \\
 19 & $(3.82 \pm 0.47) \times 10^{19}$                   & \textsc{ephem} & \ding{51} & \citet{2012-SciNote-Fienga}              \\
 20 & $(3.94 \pm 0.69) \times 10^{19}$                   & \textsc{ephem} & \ding{51} & \citet{2013-Icarus-222-Kuchynka}         \\
 21 & $(2.43 \pm 0.19) \times 10^{19}$                   & \textsc{ephem} & \ding{51} & \citet{2013-SoSyR-47-Pitjeva}            \\
 22 & $(3.82 \pm 0.41) \times 10^{19}$                   & \textsc{ephem} & \ding{51} & \citet{2014-SciNote-Fienga}              \\
 23 & $(2.72 \pm 0.12) \times 10^{19}$                   & \textsc{defl}  & \ding{51} & \citet{2014-AA-565-Goffin}               \\
 24 & $(3.42 \pm 0.18) \times 10^{19}$                   & \textsc{defl}  & \ding{51} & \citet{2014-SoSyR-48-Kochetova}          \\
 25 & $(4.18 \pm 0.44) \times 10^{19}$                   & \textsc{ephem} & \ding{51} & \citet{2017-BDL-108-Viswanathan}         \\
 26 & $0.31_{-0.31}^{0.80} \times 10^{19}$              & \textsc{defl}  & \ding{55} & \citet{2017-Icarus-297-Siltala}          \\
 27 & $(4.38 \pm 0.24) \times 10^{19}$                   & \textsc{ephem} & \ding{51} & \citet{2017-AJ-154-Baer}                 \\
\hline
 & $(3.79 \pm 1.28) \times 10^{19}$ & \multicolumn{2}{c}{Average} \\
   \hline
  \end{tabular}
\end{center}
\end{table*}

\end{appendix}
\end{document}